%% file: main.tex
\documentclass[lettersize,journal]{IEEEtran}
\usepackage{amsmath,amsfonts}
\usepackage[linesnumbered,lined,ruled,commentsnumbered]{algorithm2e}
\usepackage{array}
\usepackage[caption=false,font=normalsize,labelfont=sf,textfont=sf]{subfig}
\usepackage{textcomp}
\usepackage{stfloats}
\usepackage{url}    
\usepackage{verbatim}
\usepackage{graphicx}
\usepackage{cite}
\usepackage{algpseudocode}
\usepackage{xcolor}    
\usepackage{tikz}      
\usepackage{tipa}     
\usepackage{tcolorbox} 
\tcbuselibrary{most}  
\usepackage{booktabs}
\usepackage{multirow}
\usepackage{titlesec}
\usepackage{orcidlink}
\usepackage{circledsteps}

\newcommand{\reviseA}[1]{\textcolor{black}{#1}}
\newcommand{\reviseB}[1]{\textcolor{black}{#1}}
\newcommand{\reviseC}[1]{\textcolor{black}{#1}}

\titleclass{\subsubsubsection}{straight}[\subsection]
\newcounter{subsubsubsection}[subsubsection]
\titlespacing*{\subsubsubsection}{0pt}{3.25ex plus 1ex minus .2ex}{1.5ex plus .2ex}
\titleformat{\subsubsubsection}{\normalfont\normalsize\bfseries}{\thesubsubsubsection}{1em}{}
\makeatletter
\renewcommand\thesubsubsubsection{\thesubsubsection.\@arabic\c@subsubsubsection}
\renewcommand\p@subsubsubsection{\thesubsubsection.}
\makeatother
\begin{document}
\title{Effective User-defined Keyword Spotting with Dual-stage Matching, Multi-modal Enrollment, and Continual Adaptation}
\author{
Zhiqi Ai\orcidlink{0009-0005-1034-9972},~\IEEEmembership{Graduate Student Member,~IEEE}, 
Han Cheng\orcidlink{0009-0007-8540-3990},
Shiyi Mu\orcidlink{0000-0002-5393-0258}, 
Xinnuo Li\orcidlink{0009-0003-9600-868X}\\
Yongjin Zhou\orcidlink{0000-0002-7966-2992},~\IEEEmembership{Member,~IEEE}, and
Shugong Xu\orcidlink{0000-0003-1905-6269},~\IEEEmembership{Fellow,~IEEE}

\thanks{This work was supported in part by the 6G Science and Technology Innovation and Future Industry Cultivation Special Project of Shanghai Municipal Science and Technology Commission under Grant 24DP1501001 and in part by the National High Quality Program under Grant TC220H07D, and in part by the Xi'an Jiaotong-Liverpool University under Grant for ILAI. This work extends our IEEE ICASSP paper \cite{ds-kws} with new methods, a complete system, and extensive experiments, further demonstrating the robustness and versatility of the proposed approach across diverse keyword spotting tasks. \textit{(Corresponding authors: Yongjin Zhou and Shugong Xu.)}}%

\thanks{Zhiqi Ai, Han Cheng, Shiyi Mu, and Yongjin Zhou are with Shanghai University,
Shanghai 200444, China (e-mail: aizhiqi-work@shu.edu.cn; yjzhou@shu.edu.cn).}
\thanks{Xinnuo Li is with New York University, New York, NY 10012 USA}
\thanks{Shugong Xu is with Xi’an Jiaotong-Liverpool University, Suzhou 215000,
China (e-mail: shugong.xu@xjtlu.edu.cn).}
}
\markboth{Journal of \LaTeX\ Class Files, December~2025}%
{Shell \MakeLowercase{\textit{et al.}}: A Sample Article Using IEEEtran.cls for IEEE Journals}

\maketitle
\input{sections/0_abstart}

\input{sections/1_introduction}
\input{sections/2_related}
\input{sections/3_method}

\input{sections/4_experiments}
\input{sections/5_results}

\input{sections/6_conclusion}

\bibliographystyle{IEEEtran}
\bibliography{ref} 

\vfill
\end{document}

%% file: sections/0_abstart.tex
\begin{abstract}
User-defined keyword spotting (KWS) is crucial for personalized voice interaction, yet existing methods face several challenges:
(1) insufficient discriminability among confusable words,
(2) performance inconsistency across speakers with varying pronunciations, and
(3) high data cost to ensure reliable wake-word performance.
In this paper, we introduce \textbf{DMA-KWS}, an efficient and robust framework for user-defined keyword spotting.
First, it adopts a dual-stage matching pipeline: CTC decoding with streaming phoneme search to locate candidate segments, followed by QbyT with a phoneme matcher for fine-grained verification, enabling it to better distinguish confusable words.
Next, multi-modal enrollment fuses user-specific speech with text embeddings to further improve accuracy for registered users.
Finally, a parameter-efficient continual adaptation mechanism performs lightweight updates using synthetic and real data.
Extensive experiments demonstrate the superior performance of DMA-KWS. On the LibriPhrase Hard subset, it achieves 97.85\% AUC and 6.13\% EER, reaching state-of-the-art performance. In speaker-dependent settings, DMA-KWS consistently outperforms text-only enrollment, demonstrating significant performance gains. Moreover, the proposed parameter-efficient fine-tuning mechanism adapts DMA-KWS with only 187k updated parameters, further enhancing KWS performance while ensuring suitability for on-device deployment.
\end{abstract}


%% file: sections/1_introduction.tex
\section{Introduction}
\label{sec:intro}

\IEEEPARstart{K}{eyword} spotting (KWS) systems are typically trained on extensive corpora of predefined keywords such as “Ok Google” and “Hey Siri”~\cite{Hoy2018AlexaSC, kws_survery}, whereas customizing new wake words often entails costly data collection and retraining. With the widespread adoption of smart devices and conversational terminals (e.g., “MI XiaoAi”~\cite{shan18_interspeech}), the demand for personalized voice interaction has been rapidly increasing, driving research toward user-defined keyword spotting (UDKWS)~\cite{baseline_cmcd, qbye_lg1}. UDKWS aims to detect novel user-defined keywords from only a few enrollment examples, providing a more flexible and efficient solution for personalized voice trigger. Nevertheless, compared with predefined systems trained on large-scale corpora, current UDKWS approaches still exhibit a noticeable performance gap~\cite{wang21ea_interspeech, baseline_phonmatchnet, ai24_interspeech}.

Previous keyword spotting relied on large vocabulary continuous speech recognition (LVCSR) systems, which achieved high accuracy on predefined keywords but performed poorly on out-of-vocabulary words~\cite{motlicek2012improving, chen2013quantifying, panchapagesan16_interspeech}. Sequence-to-sequence (seq2seq) automatic speech recognition (ASR) models (e.g., Whisper~\cite{baseline_whisper}) can also be applied to keyword spotting (KWS) through various decoding algorithms, such as WFST-based decoding graphs~\cite{catt-kws, panchapagesan16_interspeech}, greedy search~\cite{zhang2022wenetspeech}, and beam search~\cite{8268974, lugosch2018donut}. However, their large model size and high computational cost make them impractical for always-on KWS on edge devices. Recently, lightweight phoneme-level ASR methods have emerged, combining phoneme detection with streaming decoding over short audio chunks to enable fast and efficient keyword recognition~\cite{cdc_kws, tdt_kw, mfa-kws}. Compared with LVCSR and word-level seq2seq ASR, phoneme-level models are more compact and can leverage large-scale ASR datasets to obtain robust pretrained speech representations. After full-shot fine-tuning, they achieve excellent KWS performance~\cite{mfa-kws}.

Lightweight query-by-example (QbyE) methods enable low-power and low-latency UDKWS by comparing registered features with query audio~\cite{wang21ea_interspeech, Qcomm}. Depending on the registration modality, they can be query-by-audio (QbyA), query-by-text (QbyT), or combined audio-and-text (QbyAT). QbyA methods extract acoustic features from registered and query audio and perform similarity-based matching, with various embedding strategies proposed to improve performance and reduce computational cost~\cite{baseline_triplet, kirandevraj2022generalized, awe3, pvtc2020}. QbyT methods leverage text-based registration to achieve more stable and robust performance, with recent work further enhancing encoder architectures, matching schemes, and pretraining strategies~\cite{baseline_cmcd, baseline_triplet, baseline_emkws, baseline_phonmatchnet, baseline_adakws, baseline_ced, ai24_interspeech, plcl, baseline_clad, SLiCK, baseline_adml, kim25d_interspeech, baseline_tts, jin24d_interspeech, baseline_ncc}. Multimodal QbyAT approaches combine the strengths of both modalities to achieve state-of-the-art results on datasets such as LibriPhrase~\cite{ai24_interspeech, plcl}. Despite their efficiency and flexibility, QbyE methods often rely on pre-segmented pairs and sliding-window detection during inference, which can lead to a mismatch with the training stage.

Despite recent progress, UDKWS still faces several critical challenges before deployment:
(1) Limited zero-shot capability: Existing models perform poorly in zero-shot scenarios, particularly when recognizing newly enrolled keywords and distinguishing confusable negative samples~\cite{baseline_cmcd, baseline_emkws, kim25d_interspeech}. Moreover, speaker accents lead to significant performance differences for the same keyword, pronunciations of the same keyword often exhibit strong speaker-dependent variations~\cite{plcl, ai24_interspeech}.
(2) Lack of efficient fine-tuning mechanisms: Most existing methods primarily focus on the pretraining stage~\cite{baseline_cmcd, ai24_interspeech, plcl, kim25d_interspeech} and lack effective continual adaptation strategies. Even after pretraining, these models still require substantial target-keyword data and incur high fine-tuning costs during the initial customization phase~\cite{baseline_adakws, seo2021wav2kws, wekws, mfa-kws}, which limits their ability to rapidly adapt in deployment scenarios.

To address these limitations, we propose DMA-KWS, an efficient UDKWS framework that integrates dual-stage matching, multi-modal enrollment, and continual adaptation mechanism. First, DMA-KWS performs a CTC-based streaming phoneme search to identify candidate segments, followed by a QbyT-based phoneme matcher for fine-grained verification. This text-only enrollment enables keyword customization for any user, and the ability to distinguish confusable keywords is further enhanced through a dual data scaling strategy, which expands both the ASR corpus and the keyword anchor set. Second, DMA-KWS extends to the speaker-dependent scenario, where multi-modal enrollment leverages the registered user’s reference audio. A Multi-modal Alignment Module fuses the speaker’s speech features with the keyword text for effective multi-modal registration, utilizing accented pronunciations in the registered audio to improve recognition accuracy for registered users. Finally, a parameter-efficient continual adaptation mechanism in DMA-KWS allows rapid fine-tuning on synthesized examples and lightweight updates with real wake-up data, achieving high recall while maintaining very low false alarm rates. Our main contributions are:

\begin{itemize}
    \item We propose DMA-KWS, an efficient framework that integrates dual-stage matching, multi-modal enrollment, and continual adaptation, providing strong zero-shot capability, effective distinction of confusable keywords, and rapid adaptation to newly enrolled keywords.
    \item We introduce a multi-modal alignment module that integrates keyword text with the registered user’s reference audio, enabling speaker-dependent keyword spotting and improving recognition accuracy for registered users.
    \item We develop a parameter-efficient continual adaptation mechanism that performs lightweight updates using synthesized and real wake-up data, enhancing recognition performance for target keywords.
    \item Extensive experiments on multiple datasets demonstrate that DMA-KWS achieves state-of-the-art performance, exhibiting strong zero-shot capability in both speaker-independent and speaker-dependent scenarios. The continual adaptation mechanism enables rapid customization of newly enrolled keywords with lightweight parameter updates, further improving KWS performance.
    \item Open-source research on user-defined keyword spotting remains limited. To advance the development of KWS, we release the training and fine-tuning code for the DMA-KWS model on GitHub\footnote{https://github.com/aizhiqi-work/DMA-KWS}.
\end{itemize}

%% file: sections/2_related.tex
\section{Related Work}
\label{sec:related}

\subsection{Predefined and User-defined Keyword Spotting}

Predefined keyword spotting (KWS) typically relies on large-scale labeled datasets~\cite{kws_survery, wekws}. For example, wake word data for MI XiaoAi can reach 1.7k hours~\cite{shan18_interspeech}, while Amazon Alexa data can reach 470 hours~\cite{8999090}. 
Negative samples are also included to form a binary classification (BCE) task. Research mainly focuses on improving detection accuracy and efficiency, including designing high-performance deep neural networks~\cite{wekws, con1}, model pruning and quantization~\cite{kws_survery}, and metric learning~\cite{huh2021metric, protoloss-kws} to optimize performance while reducing inference cost. \reviseA{User-defined keyword spotting (UDKWS) targets personalized voice interaction and usually pretrains models on large-scale ASR corpora to obtain strong feature representations~\cite{zhang2022wenetspeech, mfa-kws, kim25d_interspeech, baseline_adakws}, followed by fine-tuning with a sufficient amount of user-provided data.} Although this approach enables rapid adaptation to new keywords, its performance is still below that of predefined KWS~\cite{ai24_interspeech, mfa-kws}, particularly in low-resource scenarios. Current research primarily focuses on enhancing feature representations through pretraining to improve comparability with predefined KWS~\cite{baseline_cmcd, ai24_interspeech, kim25d_interspeech}.

\subsection{Lightweight Query-by-Example Keyword Spotting}
\label{subsec:QBYE}

Lightweight query-by-example (QbyE) methods enable low-power, low-latency user-defined keyword spotting (UDKWS) by comparing registered features with query audio~\cite{wang21ea_interspeech, Qcomm}.

\textbf{Query-by-Audio (QbyA).} QbyA is a speaker-dependent KWS system that extracts acoustic features from registered and query audio and performs similarity-based matching for UDKWS.~\cite{wang21ea_interspeech} proposed using bottleneck features (BNF) combined with dynamic time warping (DTW) for matching.~\cite{wang2021dku} further introduced phoneme posterior probabilities (PPP) as a second-stage feature matcher to improve performance. To reduce the computational cost of DTW,~\cite{awe0, baseline_triplet, awe3, hu2021acoustic} proposed using acoustic word embeddings (AWE) for similarity matching, achieving high accuracy while significantly lowering computation.

\textbf{Query-by-Text (QbyT).} QbyT is a speaker-independent user-defined keyword spotting system that constructs keyword representations aligned with speech using text inputs, such as phoneme sequences~\cite{baseline_cmcd, baseline_emkws, baseline_phonmatchnet, baseline_adakws, baseline_ced, ai24_interspeech, plcl, baseline_clad, SLiCK, baseline_adml, kim25d_interspeech, baseline_tts, jin24d_interspeech, baseline_ncc, 10633820}.~\cite{baseline_cmcd} employs attention-based cross-modal matching and sequence alignment losses for end-to-end text-to-audio alignment.~\cite{baseline_emkws} uses a DSP module for cross-modal alignment combined with DistilBERT~\cite{sanh2019distilbert} to extract word-level features, while~\cite{baseline_phonmatchnet} introduces phoneme-level auxiliary losses to improve alignment precision.~\cite{baseline_clad} leverages contrastive learning for pretraining, and~\cite{SLiCK, baseline_adml, 10633820} further enhance performance through sequence modeling, adversarial losses, and improved cross-modal matching strategies. To handle confusable keywords,~\cite{baseline_ced} augments mini-batches with hard examples, while~\cite{ai24_interspeech} employs LLMs and TTS to generate synthetic difficult samples. Additionally,~\cite{baseline_ced, baseline_adakws, kim25d_interspeech, plcl} improve model representation through large-scale CTC pretraining. In general, the QbyT text registration branch relies on phoneme-based encoders, including DistilBERT~\cite{sanh2019distilbert}, G2P embeddings~\cite{baseline_phonmatchnet}, or TTS-based text encoders~\cite{baseline_tts, baseline_ncc}, to enhance textual feature representation.

\textbf{Multimodal Query-by-Audio\&Text (QbyAT).} QbyAT integrates complementary information from both audio and text, often leveraging audio-text alignment modules (e.g., Whisper~\cite{baseline_whisper}, XLSR-53~\cite{baevski2020wav2vec}) to further enhance robustness, achieving state-of-the-art performance on datasets such as LibriPhrase~\cite{ai24_interspeech, plcl}. Compared with single-modality methods, multimodal QbyAT fully exploits the complementary strengths of audio and text, improving keyword detection accuracy and reliability while demonstrating greater adaptability under diverse registration conditions.

Despite their flexibility and efficiency, QbyE methods often rely on pre-segmented pairs and sliding-window detection during inference, which may cause mismatches with the training stage. Additionally, QbyT methods that rely on large pretrained text encoders incur substantial computational and memory overhead.

\subsection{Leveraging Synthesis Data for Keyword Spotting}
Many studies have leveraged large language models (LLMs) and speech synthesis techniques to enhance KWS performance. Prior work has explored generating negative samples related to target keywords—such as homophones, synonyms~\cite{ai24_interspeech}, and vowel-consistent confusable words~\cite{zhu25b_interspeech}—using LLMs to improve model discriminability. In addition, recent text-to-speech (TTS) systems have been widely adopted for training KWS models by combining synthetic and real data~\cite{park24_syndata4genai, lee2023fully, zhu24_syndata4genai}. For instance,~\cite{zhu24_syndata4genai} employed 7.5 million synthesized positive samples to improve model performance;~\cite{park24_syndata4genai} mixed varying ratios of TTS and real data, and used interpolation analysis to estimate the amount of real data required to achieve specific performance targets; while~\cite{lee2023fully} utilized a pseudo-TTS model  trained on a large-scale unlabeled speech corpus. The pseudo-TTS model takes pseudo-phoneme sequences extracted from wav2vec2.0~\cite{baevski2020wav2vec} and reference speech as inputs, generating utterances with diverse speaker characteristics and prosody, thereby enabling fully unsupervised few-shot KWS training. These studies show that synthetic data is crucial for fast model optimization.

%% file: sections/3_method.tex
\section{Method}

\begin{figure}[!t]
\centerline{\includegraphics[width=\linewidth]{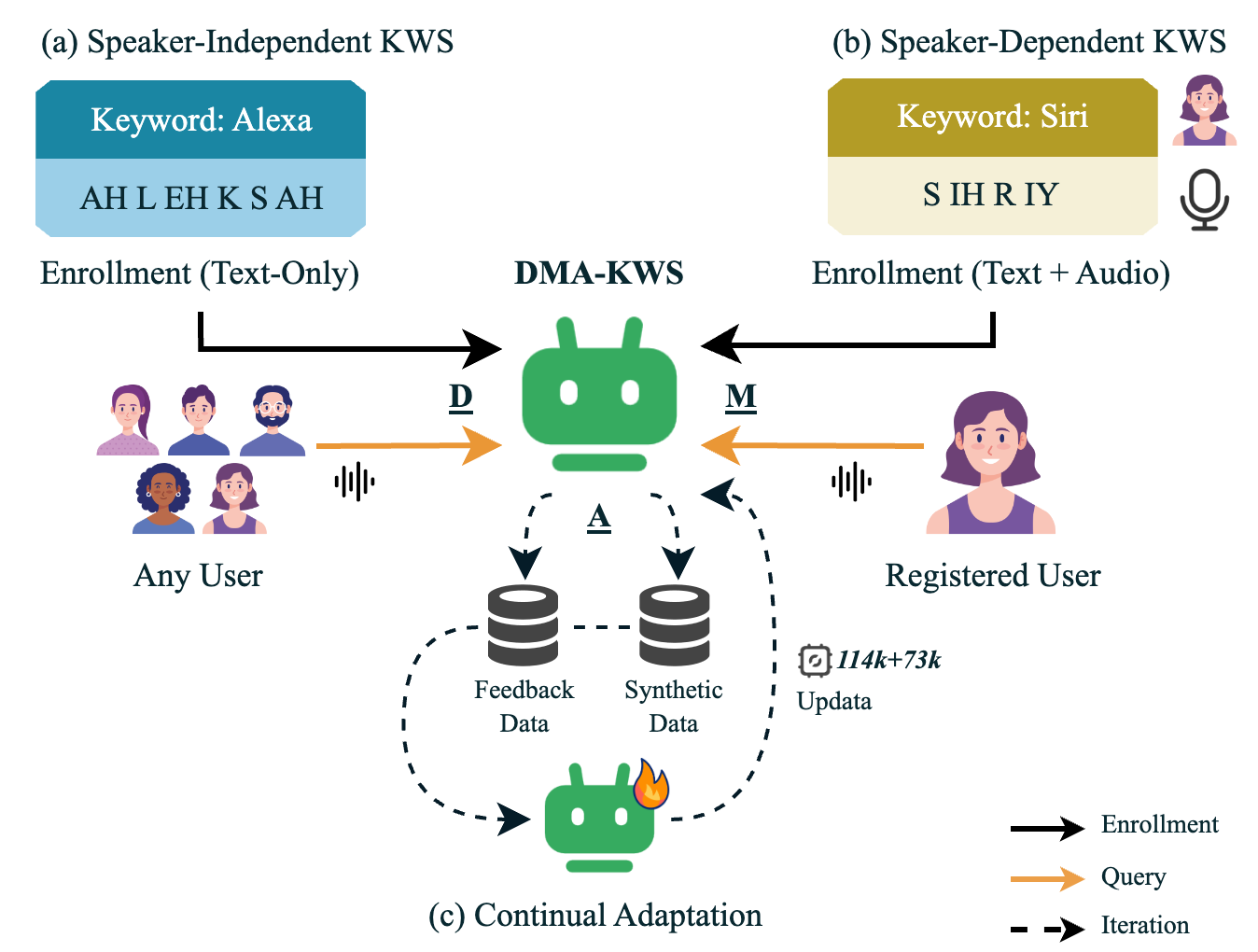}}
\caption{Schematic overview of the proposed DMA-KWS framework. (a) In the speaker-independent scenario, text-only enrollment enables keyword customization for any user; (b) In the speaker-dependent scenario, multi-modal enrollment leverages the registered user’s reference audio; (c) The continual adaptation mechanism iteratively updates the model using feedback and synthetic data.}
\label{fig.schematic-overview}
\end{figure}

\subsection{Overview}

This section provides a comprehensive overview of the proposed DMA-KWS framework, as shown in Figure \ref{fig.schematic-overview}. First, Section \ref{subsec:ctc} introduces the CTC-based keyword spotting system along with its parameter-free streaming decoding module. Next, Section \ref{subsec:d-kws} presents a QbyT-based fine-grained matching approach, which employs a phoneme-level QbyT matcher for utterance- and phoneme-level verification. Then, Section \ref{subsec:m-kws} describes the multi-modal enrollment mechanism for speaker-dependent KWS, which fuses keyword text embeddings with the registered user’s reference audio to improve performance. Finally, Section \ref{subsec:a-kws} introduces the parameter-efficient continual adaptation module, leveraging synthetic training data and online user feedback to enhance system performance.

\subsection{CTC-based Keyword Spotting System}
\label{subsec:ctc}

Inspired by\cite{baseline_ced, baseline_emkws, ai24_interspeech, kim25d_interspeech}, we adopt the Conformer~\cite{conformer} as the audio encoder. Given an input speech sequence $F_a = (f_1, \dots, f_L)$, the encoder outputs audio embeddings:
\begin{equation}
E_a = \text{Encoder}(F_a) = (\mathbf{e}_1, \dots, \mathbf{e}_T), \quad \mathbf{e}_t \in \mathbb{R}^d,
\end{equation}
where $L$ and $T$ are the lengths of the input and output sequences, respectively. The embeddings $E_a$ are mapped by the CTC layer to the posterior distribution $\mathbf{C}$ over the label set $\mathcal{A}$ (70 phonemes plus the \textit{blank} token, $|\mathcal{A}|=71$):
\begin{equation}
\mathbf{c} = \text{CTC}(E_a) = (\mathbf{c}_1, \dots, \mathbf{c}_T), \quad \mathbf{c}_t \in [0, 1]^{|\mathcal{A}|}.
\end{equation}

For the keyword text input, the sequence is converted to 70 phonemes via G2P~\cite{ai24_interspeech, baseline_cmcd, mfa-kws} and mapped to the target label sequence $\mathbf{p}$:
\begin{equation}
\label{eq:g2p}
\mathbf{p} = \text{CharTokenizer}(\text{G2P}(text)) \in \mathbb{Z}^{|\mathbf{p}|}.
\end{equation}

The model is trained with the standard CTC loss, which aims to maximize the probability of the target label sequence $P(\mathbf{p}|\mathbf{c})$:
\begin{equation}
\mathcal{L}_\text{CTC}(\mathbf{c}, \mathbf{p}) = - \log P(\mathbf{p}|\mathbf{c}) = - \log \sum_{\pi \in \mathcal{B}^{-1}(\mathbf{p})} P(\pi|\mathbf{c}).
\end{equation}
Here, $\mathcal{B}^{-1}(\mathbf{p})$ is the set of all valid label alignments (paths) $\pi$ of length $T$ that collapse to $\mathbf{p}$.

Previous studies execute frame-synchronous forced alignment via a streaming CTC decoding algorithm~\cite{mfa-kws}, which tracks the most probable phoneme paths from the posterior outputs with a complexity of $\mathcal{O}(T \times U)$. Building on this, we adopt a standard streaming decoding approach for CTC-based UDKWS to efficiently calculate the frame-level scores. The full decoding procedure is detailed in Algorithm~\ref{algo:streaming-ctc}.

\begin{algorithm}[t]
\caption{CTC Streaming Decoding}
\label{algo:streaming-ctc}
\KwIn{CTC posterior matrix $\mathbf{P} = [p_1, \dots, p_T] \in \mathbb{R}^{T \times V}$, target phoneme sequence $\mathbf{w} = [w_1, \dots, w_U]$.}
\KwOut{Frame-level score sequence $\text{Score}[1:T]$.}
\BlankLine
Insert blank tokens to obtain $\tilde{\mathbf{w}} = [\phi, w_1, \phi, \dots, \phi, w_U, \phi]$, where $\tilde{U} = 2U + 1$. \\
Initialize forward probabilities: $\delta(1,1) = \delta(1,2) = 1$. 
\BlankLine
\For{$t = 2$ \KwTo $T$}{
    \For{$\tilde{u} = 1$ \KwTo $\tilde{U}$}{
        \uIf{$\tilde{w}_{\tilde{u}} = \phi$}{
            $\delta(t,\tilde{u}) = p_t(\phi) \cdot 
            \max\{\delta(t-1,\tilde{u}-1), \delta(t-1,\tilde{u})\}$;
        }
        \Else{
            $\delta(t,\tilde{u}) = p_t(\tilde{w}_{\tilde{u}}) \cdot 
            \max\{\delta(t-1,\tilde{u}-2), \delta(t-1,\tilde{u}-1), \delta(t-1,\tilde{u})\}$;
        }
    }
    $\text{Score}[t] = \max\{\delta(t,\tilde{U}-1), \delta(t,\tilde{U})\}$;
}
\Return $\text{Score}[1:T]$.
\end{algorithm}

\subsection{Dual-stage Matching for Speaker-independent KWS}
\label{subsec:d-kws}

\begin{figure*}[!t]
\centerline{\includegraphics[width=\linewidth]{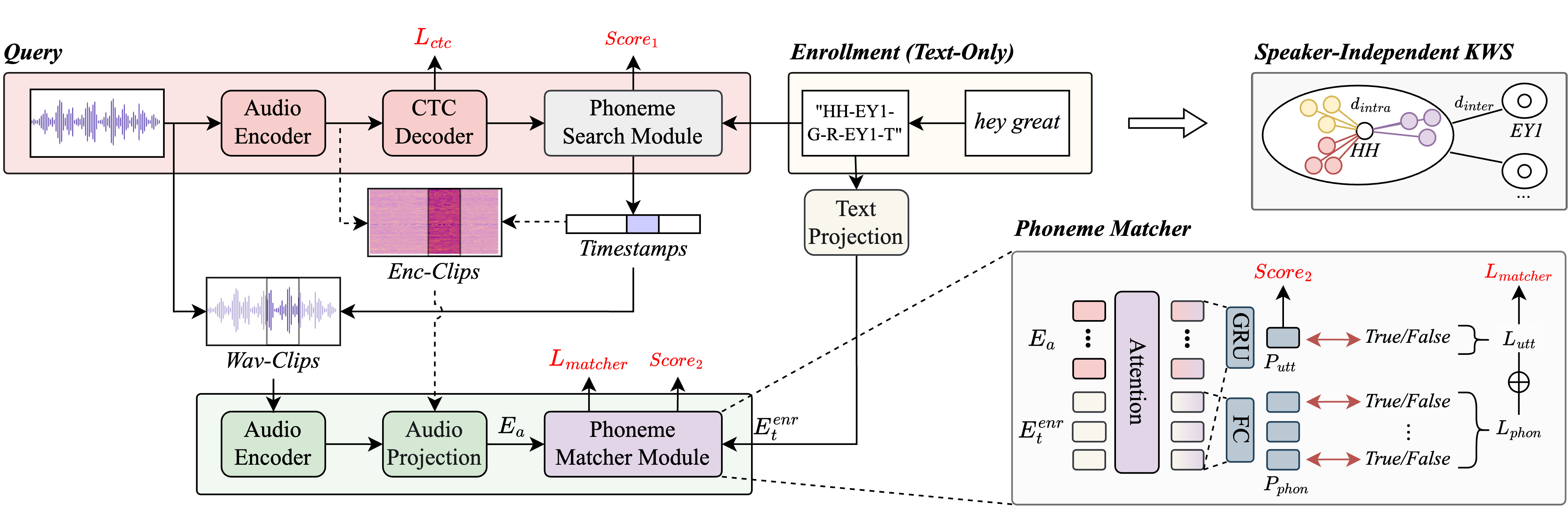}}
\caption{Overview of the speaker-independent user-defined KWS system with the proposed dual-stage matching architecture. The query audio is first processed by a CTC-based phoneme search to produce candidate segments and the first-stage score $S_1$, followed by a QbyT-based phoneme matcher that generates the second-stage score $S_2$. Text-only enrollment enables flexible customization for any user.}
\label{fig:ds-kws}
\end{figure*}

Figure \ref{fig:ds-kws} illustrates the proposed dual-stage matching strategy. As discussed in Section \ref{subsec:ctc}, the first stage performs coarse-grained keyword detection using a CTC-based branch with parameter-free streaming decoding. Its advantage is leveraging large-scale ASR data to learn more generalizable representations. However, its coarse matching limits discriminability for confusable keyword pairs. To further enhance fine-grained discrimination, the second stage introduces a QbyT-based phoneme matcher, which conducts both utterance-level and phoneme-level similarity verification on the candidate segments. Together, these two stages form a coarse-to-fine dual-stage matching architecture for robust speaker-independent keyword spotting.

The phoneme matcher determines whether an input embedding $E_a$ corresponds to a target keyword, given its enrollment prototype $E_t^\text{enr}$. Formally, this can be expressed as estimating the binary probability
\begin{equation}
\label{eq:p1}
P\big(y \,\big|\, E_a \,,\, E_t^\text{enr}\big),
\end{equation}
where $y\in \{0,1\}$ indicates the presence ($1$) or absence ($0$) of the keyword, and the phoneme sequence $\mathbf{p}$ is mapped to $E_t^\text{enr}$ through a learnable phoneme-to-embedding mapping layer.

In practice, this probability estimation can be interpreted in terms of distances between each audio embedding $e_x$ and phoneme prototypes: the embedding should be closer to the target prototype $p_\text{target}$ than to any non-target prototype $p_\text{non\_target}$:
\begin{equation}
\label{eq:p2}
\begin{aligned}
    \underbrace{d(e_x, p_\text{target})}_{d_\text{intra}} \ll \underbrace{d(e_x, p_\text{non\_target})}_{d_\text{inter}}, \quad \forall x.
\end{aligned}
\end{equation}

In our proposed dual-stage matching strategy, two input modes are supported:
\begin{itemize}
    \item \textbf{\Circled{1} Enc-Clips (Frozen):} Audio features are cropped from the CTC branch outputs based on predicted timestamps and projected through a lightweight linear layer.
    \item \textbf{\Circled{2} Wav-Clips (Trainable):} Raw audio is cropped using the same timestamps and re-encoded with a fine-tuned audio encoder for more discriminative features.
\end{itemize}

Audio and phoneme embeddings ($E_a$ and $E_t^\text{enr}$) are concatenated and passed through a self-attention layer, then processed by a lightweight discriminator to produce phoneme- and utterance-level similarity scores, $P_\text{phon}$ and $P_\text{utt}$, respectively (Figure~\ref{fig:ds-kws}). The utterance-level score $P_\text{utt}$ serves as the Stage-2 score $Score_2$. The matcher is trained with a joint loss:
\begin{equation}
\label{eq:loss1}
\mathcal{L}_{\text{matcher}} = \mathcal{L}_{\text{utt}} + \mathcal{L}_{\text{phon}},
\end{equation}
where $\mathcal{L}_{\text{utt}}$ is a BCE loss on the utterance-level prediction and $\mathcal{L}_{\text{phon}}$ is a sequence BCE loss on phoneme-level predictions. The overall training objective combines both stages:
\begin{equation}
\mathcal{L}_{\text{total}} = \mathcal{L}_{\text{CTC}} + \mathcal{L}_{\text{matcher}}.
\end{equation}

\subsection{Multi-modal Enrollment for Speaker-dependent KWS}
\label{subsec:m-kws}

\begin{figure*}[!t]
\centerline{\includegraphics[width=0.95\linewidth]{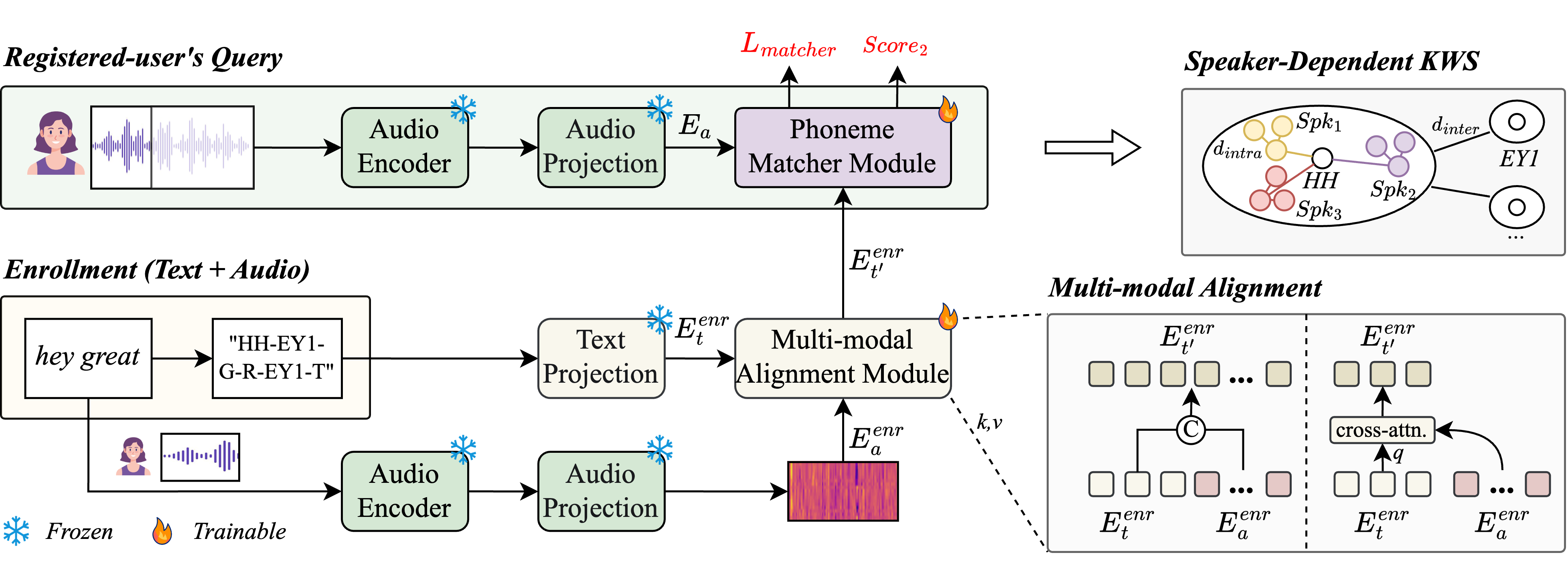}}
\caption{Overview of the speaker-dependent user-defined KWS system with the proposed multi-modal enrollment architecture. The enrollment leverages the registered user's reference audio via the Multi-modal Alignment Module (MAM). Training updates only the Phoneme Matcher Module, which computes the matching score $Score_2$, and the MAM, while all other model parameters remain frozen.}
\label{fig:ds-kws-at}
\end{figure*}

Figure \ref{fig:ds-kws-at} illustrates the proposed multi-modal enrollment strategy. Building on the audio encoder outputs described in Section \ref{subsec:d-kws}, we design a multi-modal alignment module (MAM) to fuse audio and text information. The phoneme matcher then performs utterance-level and phoneme-level similarity verification on candidate segments. By integrating the registered user’s speech features with the keyword text, effectively leveraging accented pronunciations present in the audio, this approach enables speaker-dependent keyword spotting.

The Phoneme Matcher determines whether an input embedding $E_a$ corresponds to a target keyword, given its enrollment embeddings $E_t^\text{enr}$ (text) and $E_a^\text{enr}$ (audio). Formally, this can be expressed as estimating the binary probability
\begin{equation}
P\big(y \,\big|\, E_a \,,\, (E_t^\text{enr}, E_a^\text{enr})\big),
\end{equation}
following the formulation in Equation \eqref{eq:p1}. Both $E_a$ and $E_a^\text{enr}$ are obtained using the pretrained audio encoder described in Section \ref{subsec:d-kws}, and the probability estimation follows the method outlined in Equation \eqref{eq:p2}.

These two modalities are fused via MAM to produce the unified speaker-aware prototype $E_{t'}^{enr}$. We explore two fusion strategies:
\begin{itemize}
    \item \textbf{\Circled{3} Concat (no extra parameters):} 
    Text and audio embeddings are concatenated along the temporal dimension with positional encodings, 
    $E_{t'}^{\text{enr}} = [E_t^\text{enr}; E_a^\text{enr}]$. Alignment is then handled by the phoneme matcher’s self-attention. 
    \item \textbf{\Circled{4} Cross-Attention (no extra tokens):} 
    A cross-attention layer aligns the keyword’s acoustic features with text embeddings, $E_{t'}^{\text{enr}} = \text{Cross-Attn}(E_t^\text{enr}, E_a^\text{enr}, E_a^\text{enr})$. Conditioning phoneme prototypes on the enrolled speaker. 
\end{itemize}
The phoneme matcher and MAM is trained with the same objective as in Equation~\eqref{eq:loss1}.

\begin{figure}[!t]
\centerline{\includegraphics[width=0.95\linewidth]{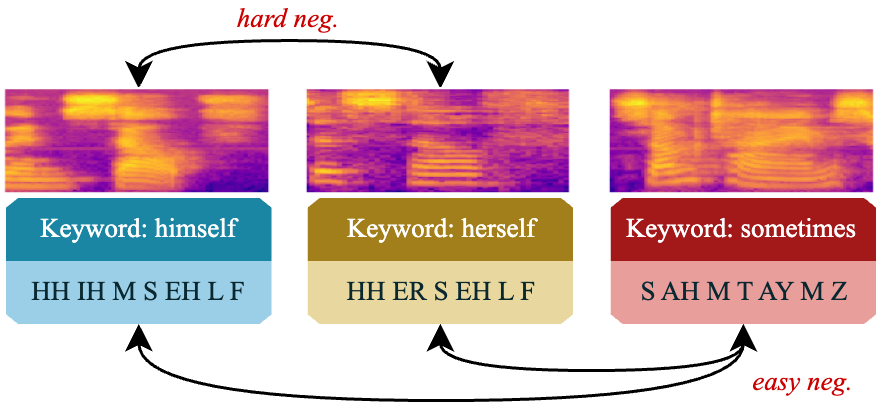}}
\caption{Illustration of sample pairs with hard and easy negatives.}
\label{fig:data-pair}
\end{figure}

Figure~\ref{fig:data-pair} shows the sample pairs for QbyE training. Hard negatives have smaller phoneme-level edit distances to positives, making them harder to discriminate. In Section~\ref{subsec:d-kws}, text-only enrollment checks text-audio consistency, while in Section~\ref{subsec:m-kws}, audio+text enrollment verifies test audio against the enrolled keyword.

\subsection{Continual Adaptation with Synthetic and Feedback Data}
\label{subsec:a-kws}

To achieve user-defined wake-word customization with low false alarm rates and minimal training cost, we adopt a parameter-efficient continual adaptation strategy based on LoRA~\cite{Hu2021LoRALA}, updating only the QKV matrices of the phoneme matcher’s attention layers (Section \ref{subsec:d-kws}) while keeping all other model parameters frozen. The adaptation process sequentially leverages synthetic and real data for fine-tuning:
\begin{itemize}
    \item \textbf{Synthetic Data:} Generate positive and confusable negative samples (optional) via TTS models (e.g., F5-TTS~\cite{Chen2024F5TTSAF}) and fine-tune the LoRA adapter to enhance discriminability.
    \item \textbf{Real Data:} Further fine-tune the LoRA adapter with real user feedback or collected recordings to refine keyword-specific performance.
\end{itemize}

LoRA updates are applied to the phoneme matcher as:
\begin{equation}
    W_\text{attn} \leftarrow W_\text{attn} + \Delta W = W_\text{attn} + A B,
\end{equation}
where \(A \in \mathbb{R}^{d \times r}\) and \(B \in \mathbb{R}^{r \times d}\) are low-rank matrices learned during adaptation, and \(r \ll d\) controls the update rank.

%% file: sections/4_experiments.tex
\section{Experimental Setup}
\subsection{Data Configuration}

We evaluate the proposed DMA-KWS framework across three key scenarios:
(1) Speaker-Independent KWS (SI-KWS): Evaluated on major benchmark datasets to assess text-only enrollment performance, including zero-shot generalization and the ability to distinguish confusable keywords.
(2) Speaker-Dependent KWS (SD-KWS): Evaluated on enrolled speakers to assess how multi-modal enrollment enhances speaker-specific keyword detection.
(3) Wake-Up Word (WUW) Tasks: Evaluated on diverse wake-word datasets and the effectiveness of the parameter-efficient continual adaptation strategy for newly registered words.

\subsubsection{ASR-Training} 
We use LibriSpeech\footnote{\url{https://openslr.org/87}}~\cite{panayotov2015librispeech} (LS) and GigaSpeech\footnote{\url{https://github.com/SpeechColab/GigaSpeech}}~\cite{chen21o_interspeech} (GS) as training corpora. For LibriSpeech, we employ the “clean” subsets \texttt{train-clean-100} (LS-100) and \texttt{train-clean-100/360} (LS-460), which contain 100 and 460 hours of high-quality read speech with transcripts, respectively.

To further expand the training data, we include the 1,000-hour subset of GigaSpeech (GS-1000), derived from the “Middle” portion of the corpus and containing approximately 910,140 audio samples. These datasets are combined to form the LS-GS-1460 training set.

\subsubsection{Phrase-Training}
We construct LibriPhrase\footnote{\url{https://github.com/gusrud1103/LibriPhrase}} (LP) following~\cite{baseline_cmcd, baseline_emkws, baseline_ced, kim25d_interspeech, baseline_adakws, ai24_interspeech, plcl}, retaining anchors with durations between 0.5–2 seconds. To study the effect of anchor-class size, the training set includes LP-100 ($\sim$12k classes) and LP-460 ($\sim$78k classes), with two sampled subsets from LP-460: LP-460-20k and LP-460-40k. GigaPhrase\footnote{\url{https://github.com/aizhiqi-work/GigaPhrase-1000}} is derived from the GigaSpeech corpus using the same MFA-based segmentation, providing a large collection of phrase-level anchors. Combining GP-1000 with LP-460 forms the LP–GP–1460 training set, containing approximately 155k phrase classes.

\subsubsection{Evaluations}
To track the performance of ASR pretraining, we use the standard LibriSpeech clean and other test sets ($LS_{clean}$ and $LS_{other}$). The evaluation metric is the phoneme-level Word Error Rate (P-WER), where all text transcripts are converted into a set of 70 phonemes (Equation~\ref{eq:g2p}).

For evaluating SI-KWS, we use the LibriPhrase test set, derived from the \texttt{train-others-500} subset of LibriSpeech. Negative examples are categorized into easy ($LP_E$) and hard ($LP_H$) subsets based on Levenshtein distances. To further assess cross-domain performance, we also evaluate on GSC and Qcomm. GSC\footnote{\url{https://huggingface.co/datasets/google/speech_commands}}~\cite{warden2018speech} contains one-second recordings of 10 English command words (“yes”, “no”, “up”, “down”, “left”, “right”, “on”, “off”, “stop”, “go”), comprising 64,720 evaluation samples and 237k anchor-pairs for detection. Qcomm\footnote{\url{https://www.qualcomm.com/developer/project/keyword-speech-dataset}}~\cite{Qcomm} includes 4,270 utterances of four keywords (“Hey Android,” “Hey Snapdragon,” “Hi Galaxy,” “Hi Lumina”) spoken by up to 50 speakers. Evaluation metrics are primarily Area Under the Receiver Operating Characteristic Curve (AUROC) and Equal Error Rate (EER).

For evaluating SD-KWS, we re-split the LibriPhrase test set by selecting positive pairs from the same speaker with the same keywords. The number of positive pairs for keyword lengths 1--4 are approximately 35k, 35k, 10k, and 1k, respectively. Negative examples are retained from the original dataset and categorized into easy ($LP^{SD}_E$) and hard ($LP^{SD}_H$) subsets. To further assess cross-domain performance, we also evaluate on $Qcomm^{SD}$~\cite{Qcomm} and AudioMNIST. AudioMNIST\footnote{\url{https://github.com/soerenab/AudioMNIST}}~\cite{AudioMNIST} consists of $\sim$30k recordings (~9.5 hours) of spoken digits (0–9) in English from 60 speakers. Evaluation metrics are primarily AUROC and EER, consistent with the SI-KWS evaluation.

For wake-up word (WUW) tasks, we evaluate primarily on the Hey-Snips dataset\footnote{\url{https://github.com/sonos/keyword-spotting-research-datasets}}
~\cite{snips}, comparing with mainstream supervised methods. Additionally, we conduct a two-stage evaluation on four supplementary wake words on the Qcomm (“Hey Android,” “Hey Snapdragon,” “Hi Galaxy,” “Hi Lumina”). On the DeepMine\footnote{\url{https://data.deepmine.ir/}}~\cite{deepmine} dataset (Persian accent), we select five words ("OK Google", "My voice is my password", "Actions speak louder than words", "Artificial intelligence is for real", "There is no such thing as a free lunch") for ASR performance evaluation and first-stage fine-tuning, with a second-stage wake-up word evaluation and fine-tuning specifically for “OK Google.” All evaluations use Recall@FAR, measuring recall at a fixed false alarm rate. Additionally, the false alarm dataset is constructed from the official Hey-Snips negative utterances by aggregating all negatives across the train, development, and test sets, yielding approximately 97 hours of audio.

\subsection{Training Details}
\textbf{Acoustic Features}
We use 80-dimensional log Mel filterbank (FBank) features with a 25 ms window, 10 ms hop, and 16 kHz sampling rate, following the standard WeNet\cite{zhang2022wenetspeech} frontend with a 0.1 dither. During CTC training, speed perturbation ({0.9, 1.0, 1.1}) and SpecAugment~\cite{park2019specaugment} (max time mask 50, max frequency mask 10) are applied. In the QbyT and SD-KWS fine-tuning stages, only clean features are used without additional augmentation.

\textbf{Model Architecture}
The DMA-KWS framework uses a lightweight 6-layer Conformer encoder with 144-dimensional embeddings, a 576-unit feed-forward module, 4 attention heads, convolution modules with kernel size 3, and relative positional self-attention. A conv2d frontend is applied, and the encoder has approximately 3.6 M parameters, shared across all stages. The QbyT branch adds a small matching head with two Transformer layers, a GRU, and fully connected layers, using text embeddings from an \texttt{nn.Embedding} module. This branch contains approximately 0.5 M trainable parameters.

\textbf{Training Details}
The first stage trains the Conformer encoder with CTC loss using WeNet~\cite{zhang2022wenetspeech}. The phoneme inventory contains 71 symbols (including the blank) generated via G2P, and separate models are trained on LS-100, LS-460, and LS-GS-1460 following all WeNet default preprocessing steps, including 16 kHz resampling, sequence filtering, sorting, and static batching. The second stage trains the phoneme matcher (QbyT branch) for 50k steps under \Circled{1} and \Circled{2} on phrase-level datasets with increasing anchor-class diversity, namely LP-100 ($\sim$12k classes), LP-460 ($\sim$78k classes), and LP–GP–1460 ($\sim$155k classes), without speed perturbation or SpecAugment. For speaker-dependent KWS (SD-KWS), the model initialized from the speaker-independent stage (\Circled{2}) is further finetuned on enrollment data for 20k steps under \Circled{3} and \Circled{4} with a learning rate of 4e-4. A parameter-efficient fine-tuning strategy based on LoRA is adopted in both stages. \reviseC{To mitigate catastrophic forgetting, the fine-tuning set is balanced with equal amounts of general ASR and phrase data.} Specifically, the first-stage encoder is fine-tuned with rank 16 using 114k learnable parameters for 5k steps at a learning rate of 4e-4 (about 30 minutes), while the second-stage matcher is fine-tuned with the same rank introducing 73k parameters, resulting in a total of 187k trainable parameters and requiring about 3 minutes. Model training is conducted on four NVIDIA RTX 4090 GPUs, whereas all fine-tuning experiments are performed on a single RTX 4090 GPU.

%% file: sections/5_results.tex
\section{results}
\input{tabs/si-libriphrase}

\subsection{Comparative Evaluation of SI-KWS}

\textbf{Baselines:} Under the SI-KWS setup, we compare against several advanced speaker-independent KWS systems, with LibriPhrase~\cite{baseline_cmcd} serving as the main benchmark (see Section~\ref{subsec:QBYE}). Our baselines include classical CMCD~\cite{baseline_cmcd}, PhonMatchNet~\cite{baseline_phonmatchnet}; ASR-pretrained EMKWS~\cite{baseline_emkws}, CED~\cite{baseline_ced}, AdaKWS~\cite{baseline_adakws}, PLCL~\cite{plcl}, SLiCK~\cite{SLiCK}, W-CTC~\cite{kim25d_interspeech}; and hard-negative-enhanced CED, PLCL, MM-KWS~\cite{ai24_interspeech}. \reviseC{Additionally, we report the performance under different pretraining conditions and data scales to demonstrate the impact of these factors on system performance.}

Table~\ref{table:lp} presents a comprehensive comparison between the proposed DMA-KWS and existing baselines on the SI-KWS task. On the in-domain LibriPhrase benchmark (Easy: $LP_E$, Hard: $LP_H$), DMA-KWS clearly outperforms all baseline systems. In particular, on the more challenging $LP_H$ subset with a higher degree of word confusability, DMA-KWS(\Circled{2}) achieves 97.85\% AUROC / 6.13\% EER, outperforming the ASR-pretrained PLCL (Whisper-based~\cite{baseline_whisper}, 40M) and also surpassing MM-KWS, which relies heavily on hard-negative augmentation. \reviseC{Even without additional data expansion from GigaSpeech or GigaPhrase, the model maintains strong performance (97.03\% AUROC / 7.97\% EER)}. Meanwhile, the frozen-encoder variant DMA-KWS(\Circled{1}) also delivers competitive results (95.77\% AUROC / 10.02\% EER). On the $LP_E$ subset, DMA-KWS(\Circled{2}) further reaches 0.45\% EER.

In out-of-domain evaluation, DMA-KWS continues to lead on both GSC and QComm (GSC: 99.21\% AUROC / 3.93\% EER; QComm: 99.90\% AUROC / 1.52\% EER), consistently outperforming all previously reported comparison systems and demonstrating strong generalization capability. Moreover, DMA-KWS(\Circled{2}) shows clear improvements over the frozen variant DMA-KWS(\Circled{1}) across both in-domain and out-of-domain benchmarks, in which the EER is reduced by 3.89\%, 2.45\%, and 0.10\% on $LP_H$, GSC, and QComm, respectively.

\subsection{Dual Data Scaling Evaluation of Dual-stage Matching}

\input{tabs/data-scaling}
\input{tabs/anchor_scaling}

\textbf{Scaling the Training Corpus:} In Stage~1, simply enlarging the ASR training corpus leads to a clear reduction in P-WER. As shown in Table~\ref{tab:dual_data_scaling}, increasing the pre-training data from 100h to 460h reduces P-WER on $LS_{clean}$ and $LS_{other}$ from 6.98\% / 18.79\% to 4.44\% / 13.39\%, respectively. Further adding GS-1000 to form LS-GS-1460 yields a continued reduction on $LS_{other}$ to 11.80\%, indicating that larger ASR pre-training data consistently improves phoneme recognition. In Stage~2, DMA-KWS(\Circled{1}) exhibits monotonic gains as data increases (100h $\rightarrow$ 460h $\rightarrow$ 1460h). LP\(_H\) AUROC improves 91.78\% $\rightarrow$ 95.33\% $\rightarrow$ 95.77\%, and EER reduces 15.34\% $\rightarrow$ 10.78\% $\rightarrow$ 10.02\%, with similar growth patterns also seen on LP\(_E\). DMA-KWS(\Circled{2}), which allows encoder fine-tuning, yields even larger improvements. Under joint scaling of Stage~1 and Stage~2, LP\(_H\) AUROC rises 93.10\% $\rightarrow$ 97.03\% $\rightarrow$ 97.85\%, while EER declines 13.71\% $\rightarrow$ 7.97\% $\rightarrow$ 6.13\%.

\textbf{Effect of Anchor Scaling in Stage~2-Only:} To further assess the role of Stage~2 data scaling, we conduct an ablation by varying the number of anchor classes, using the LS-460 ASR-pretrained model with DMA-KWS(\Circled{1}) under a frozen audio encoder. As shown in Table~\ref{tab:stage2_ablation}, performance improves consistently as the anchor set expands from 12k $\rightarrow$ 20k $\rightarrow$ 40k $\rightarrow$ 78k $\rightarrow$ 155k. On $\text{LP}_H$, EER decreases from 13.38\% to 10.65\%, while AUROC increases from 93.22\% to 95.45\%. These results show that enlarging the anchor class substantially improves the ability to distinguish confusable keywords.

\subsection{Visual Analysis of DMA-KWS Representations}

\begin{figure}[!t]
\centerline{\includegraphics[width=\linewidth]{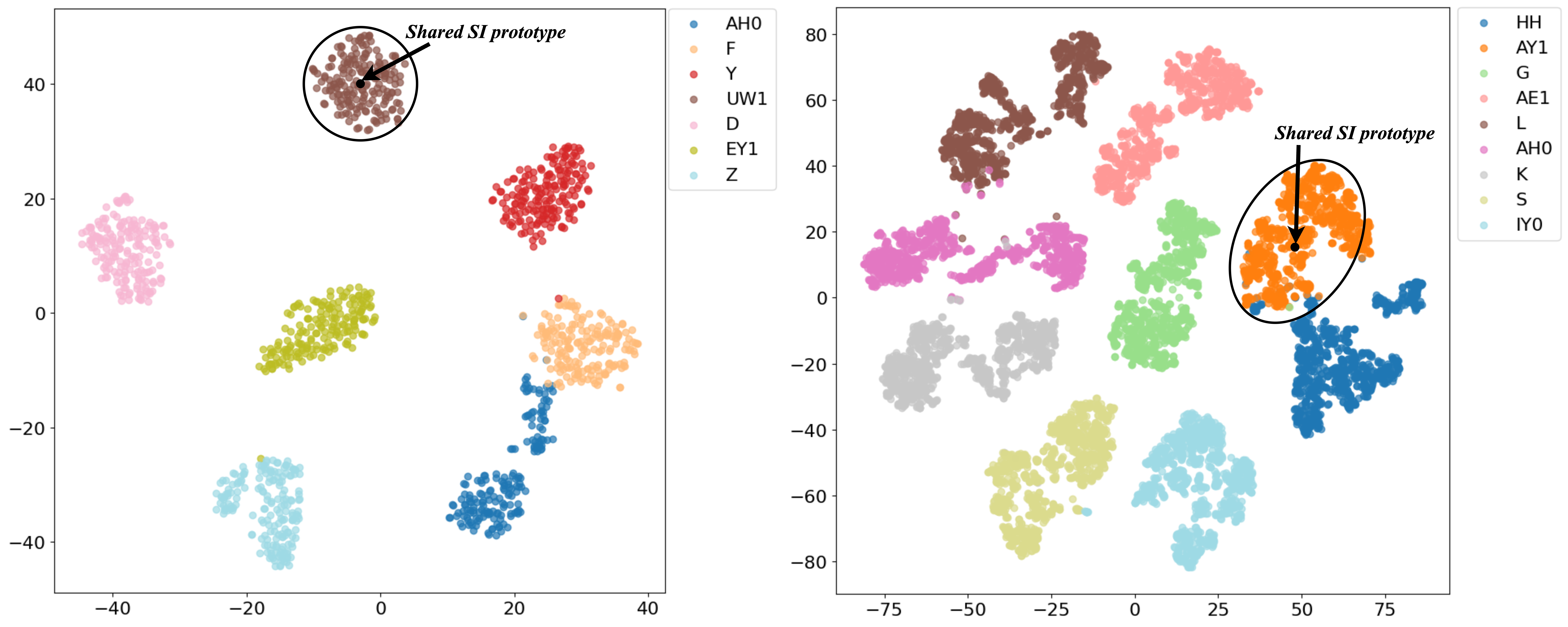}}
\caption{\reviseA{t-SNE visualization for various phonemes of DMA-KWS in the speaker-independent setting. 
The shared SI prototype is marked as black dots.}}
\label{fig.clustring_old}
\end{figure}

\begin{figure}[!t]
\centerline{\includegraphics[width=\linewidth]{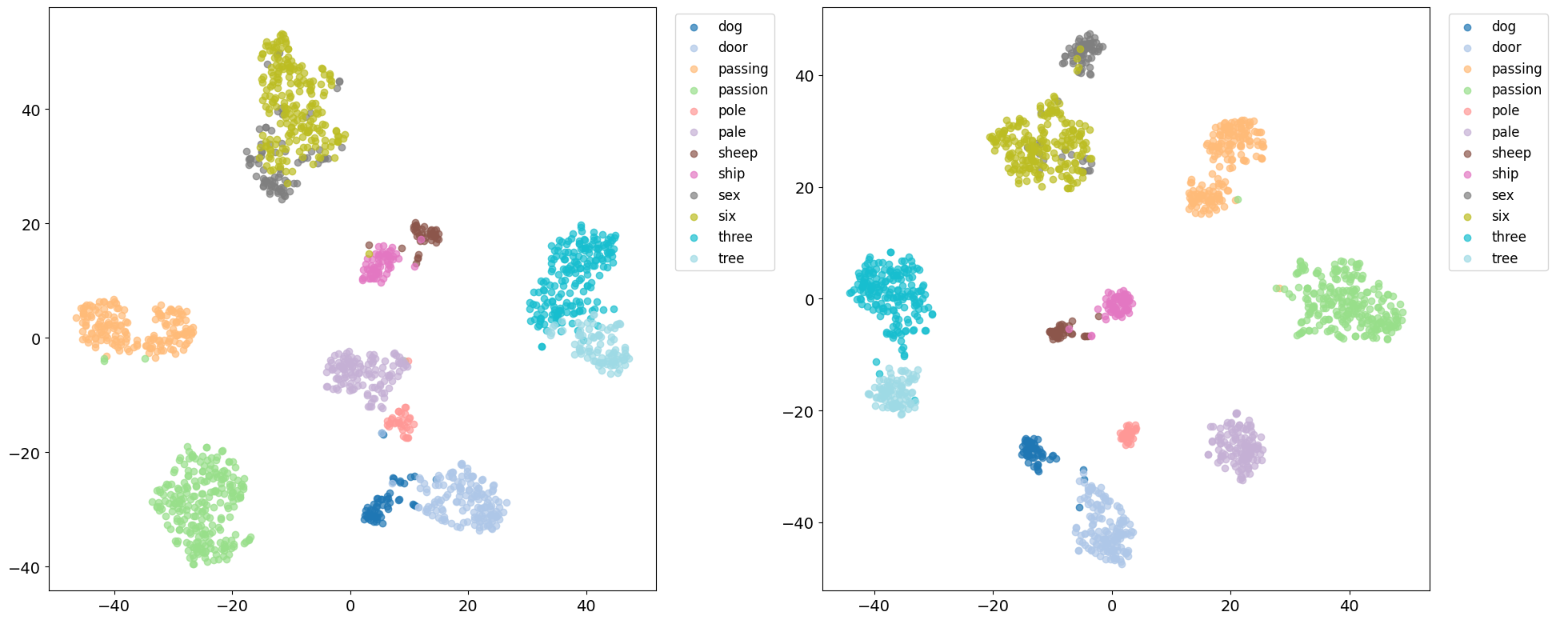}}
\caption{t-SNE visualization of challenging keywords (e.g., "sex" vs. "six"). Left: DMA-KWS(\Circled{1}); Right: DMA-KWS(\Circled{2}). The embeddings are computed using average pooling.}
\label{fig.clustring_hard}
\end{figure}

We visualize the internal representations and inference behavior of DMA-KWS in four aspects. Figure~\ref{fig.clustring_old} shows t-SNE embeddings of phoneme-level audio features from DMA-KWS(\Circled{1}), computed via average pooling over phoneme segments obtained from forced alignments. Clear separation between phonemes demonstrates that ASR pre-training produces discriminative acoustic representations. Figure~\ref{fig.clustring_hard} compares DMA-KWS(\Circled{1}) and (\Circled{2}) on six challenging keyword pairs (e.g., "sex" vs. "six"). DMA-KWS(\Circled{1}) exhibits limited discriminability for these hard examples, while DMA-KWS(\Circled{2}), after Phrase-level fine-tuning, shows significantly improved separation, highlighting the enhancement of acoustic encoding.


\begin{figure}[!b]
\centerline{\includegraphics[width=\linewidth]{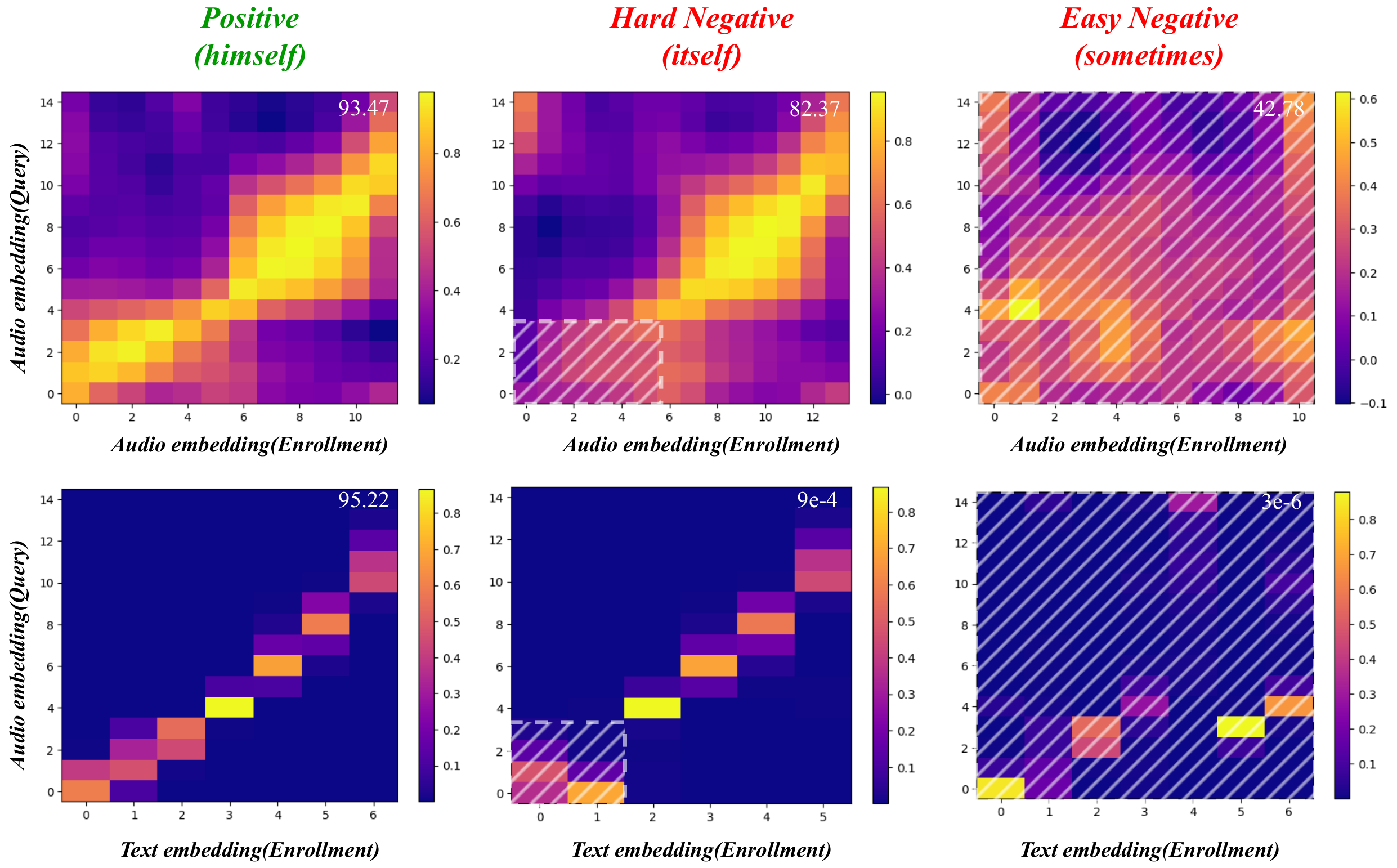}}
\caption{Heatmaps of registered and query features for DMA-KWS(\Circled{2}) showing positive examples, easy negatives, and hard negatives. Top: audio-based enrollment (cosine similarity); Bottom: text-based enrollment (attention map).}
\label{fig.visual}
\end{figure}

\begin{figure*}[!t]
\centerline{\includegraphics[width=\linewidth]{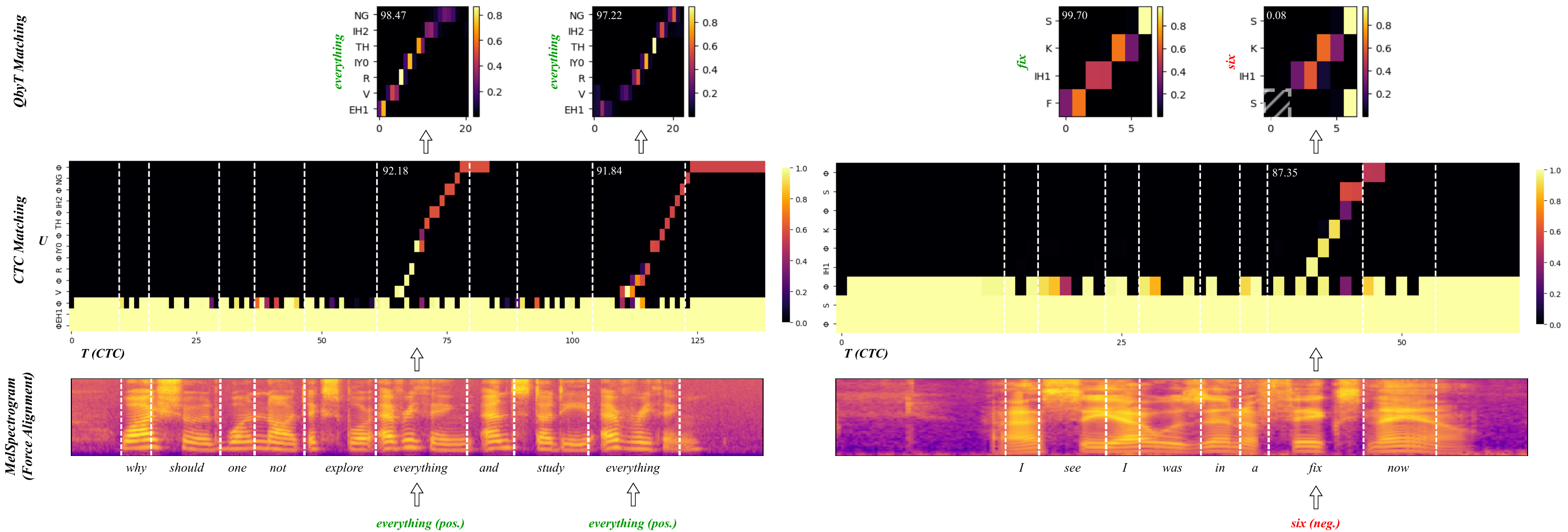}}
\caption{Heatmaps of wake-up scores at each (t, u) for the CTC branch and QbyT branch, representing a two-stage process: the utterance is first filtered by the CTC branch, and segments exceeding a threshold are re-scored by the QbyT branch. The utterance is selected from the test-clean dataset. Left: "everything" as the keyword (appears twice in the utterance); Right: "six" as the keyword (does not appear in the utterance). Vertical yellow dashed lines indicate word boundaries derived from force alignments.
}
\label{fig.ds-kws-inference}
\end{figure*}

Figure~\ref{fig.visual} presents heatmaps of registered and query features for DMA-KWS(\Circled{2}) including positive, easy negative, and hard negative examples. For positive examples, both audio- and text-based enrollment exhibit strong monotonicity (audio: 93.47, text: 95.52). For hard negatives, text-based registration shows higher discriminability, with low activations for missing phonemes (audio: 82.37, text: 9e-4), while simple negatives yield near-zero scores (audio: 42.78, text: 3e-6), indicating that text-based features provide stronger separation for challenging cases. Figure~\ref{fig.ds-kws-inference} visualizes wake-up scores at each $(t, u)$ for the two-stage process combining CTC and QbyT branches. For a positive example ("everything"), both branches produce monotonic, accurate scores. For a hard negative ("six"), the CTC branch incorrectly produces a high score, while the QbyT branch correctly assigns a low score, demonstrating its strong discriminability under challenging conditions.

\input{tabs/sd-libriphrase}

\subsection{Comparative Evaluation of SD-KWS}
\textbf{Baselines:} In the SD-KWS setting, we compare DMA-KWS with several speaker-dependent keyword spotting (KWS) systems, with the speaker-dependent LibriPhrase serving as the main benchmark (see Section~\ref{subsec:QBYE}). Our baselines include: (1) SI-KWS reference baselines: DMA-KWS(\Circled{1}), DMA-KWS(\Circled{2}), MM-KWS@T~\cite{ai24_interspeech};  (2) SD-mode systems: MM-KWS$@$AT, Whisper$@$A~\cite{baseline_whisper}, WavLM$@$A~\cite{chen2022wavlm}, DMA-KWS(\Circled{1})$@$A, and DMA-KWS(\Circled{2})$@$A\footnote{$@$T indicates discrimination using text-based enrollment; $@$A indicates discrimination using audio-encoder embeddings; $@$AT indicates multi-modal enrollment capability.}.

Table~\ref{table:sd-lp} compares the proposed DMA-KWS with existing baselines on the speaker-dependent KWS (SD-KWS) task. On the in-domain LibriPhrase benchmark (Easy: $LP^{SD}_E$, Hard: $LP^{SD}_H$), DMA-KWS(\Circled{3} and \Circled{4})$@$AT clearly outperforms all baseline systems. Notably, on the more challenging $LP^{SD}_H$ subset with higher word confusability, DMA-KWS(\Circled{4}) achieves 97.70\% AUROC and 6.58\% EER, surpassing MM-KWS$@$AT, which also adopts multi-modal enrollment but relies on a substantially larger XLSR-53~\cite{baevski2020wav2vec} audio encoder (331.8M). Moreover, compared with the text-based registration variant DMA-KWS(\Circled{2})$@$T (97.31\% AUROC / 7.21\% EER), DMA-KWS(\Circled{4}) attains clear performance improvements on $LP^{SD}_H$, further validating the significant advantage of speaker-dependent enrollment. In out-of-domain evaluations, DMA-KWS also demonstrates strong competitiveness on QComm$^{SD}$ and AudioMNIST, ranking second only to WavLM~\cite{chen2022wavlm} (97.4M-parameter model pre-trained on over 60k hours of data). Compared with the text-based registration variant DMA-KWS(\Circled{2})$@$T, DMA-KWS(\Circled{4})$@$AT maintains a clear advantage, for example reducing the EER on QComm$^{SD}$ from 1.52\% to 0.88\%.

\begin{figure}[!t]
\centerline{\includegraphics[width=\linewidth]{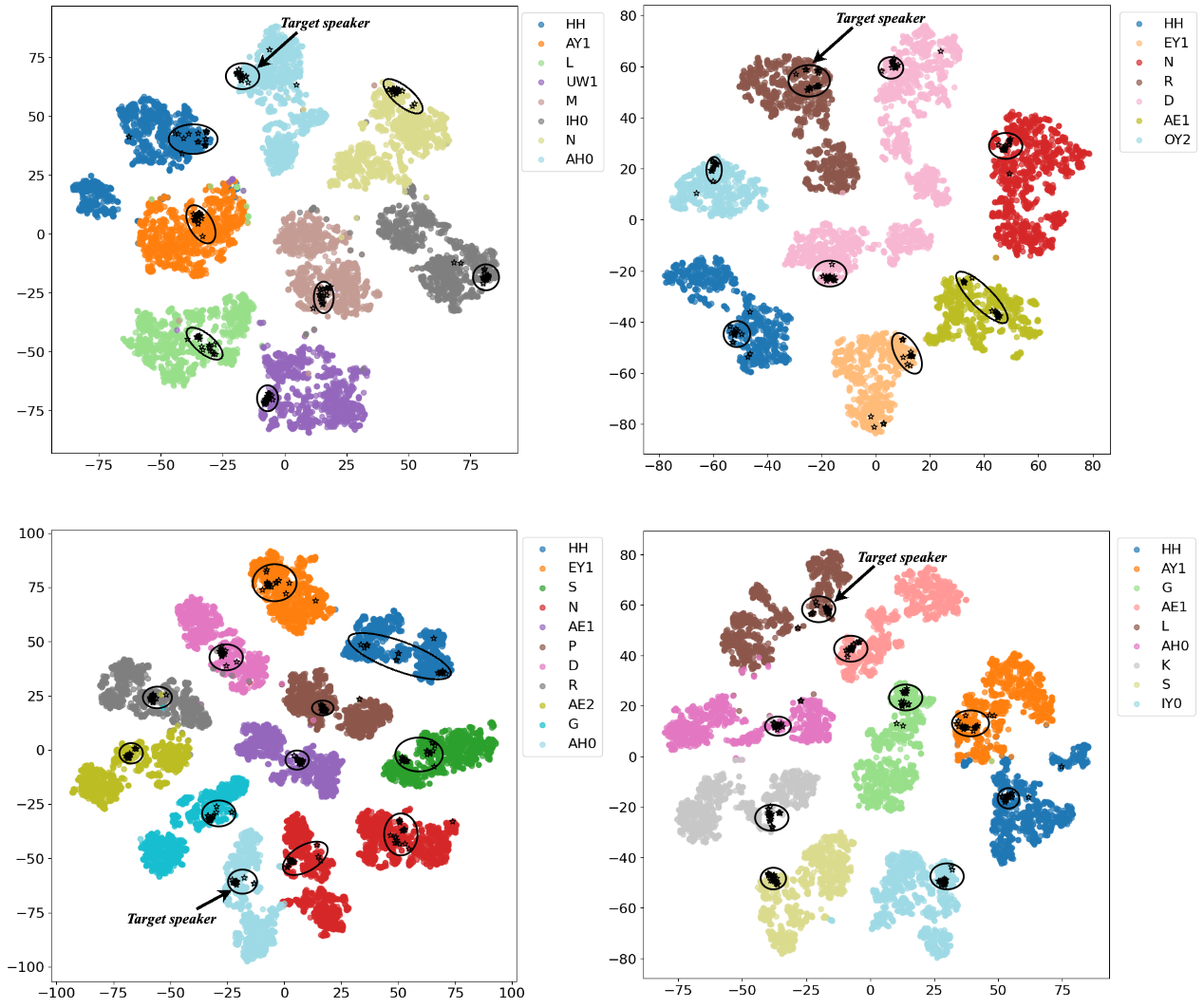}}
\caption{t-SNE visualization for various phonemes of DMA-KWS in the speaker-dependent setting. Target-speaker phoneme samples are marked in black to show discriminability.}
\label{fig.clustring_speaker}
\end{figure}

Figure~\ref{fig.clustring_speaker} shows t-SNE visualization of phoneme-level embeddings in the speaker-dependent (SD) setting, with target-speaker phonemes marked in black. In SD, both text and reference audio from the target speaker are used, whereas in SI-KWS (Figure~\ref{fig.clustring_old}) a single text prototype is shared across all speakers. Phonemes from the same speaker in SD exhibit higher correlation, forming tighter and more separable clusters, which explains the observed performance gains in SD-KWS.

\subsection{Comparative Evaluation of WUW}
\textbf{Baselines:} For wake-up word detection tasks, we primarily compare with fully supervised systems fine-tuned on the complete training data, including RIL-KWS~\cite{zhang20u_interspeech}, WaveNet~\cite{snips}, MDTC~\cite{wekws}, and the latest state-of-the-art model MFA-KWS~\cite{mfa-kws} (CTC+Transducer).

\input{tabs/snips}

\reviseA{Table~\ref{tab:snips_results} shows the zero-shot performance of DMA-KWS under different settings.} On the Hey-Snips dataset, the baseline CTC-Streaming achieves 98.06\% Recall at FAR = 0.05/h. In speaker-independent settings, DMA-KWS(\Circled{1}) reaches 98.66\%, and DMA-KWS(\Circled{2}) further increases to 99.45\%. In speaker-dependent settings, DMA-KWS(\Circled{3}) and DMA-KWS(\Circled{4}) achieve 99.64\% and 99.72\%, respectively.

Table~\ref{tab:customs_results} shows the recall and average performance for additional custom wake words at FAR = 0.05/h. Compared with the CTC-Streaming baseline, DMA-KWS(\Circled{1}) achieves an average recall of 94.50\%, DMA-KWS(\Circled{2}) reaches 96.86\%, and DMA-KWS(\Circled{3}) and DMA-KWS(\Circled{4}) achieve 98.01\% and 98.70\%, respectively. These results demonstrate that DMA-KWS performs very well on multiple custom wake words under zero-shot settings.

\input{tabs/few-shot-kws} 

\subsection{Parameter-efficient Fine-tuning for Target Wake Words}
We first evaluate the Stage-1 CTC model on the DeepMine dataset, as summarized in Table~\ref{tab:few-shot}. Zero-shot (ZS) provides a baseline, and we observe a notable performance gap compared with general datasets (P-WER = 11.80\% on $LS_{other}$). We then perform ASR fine-tuning using different strategies and data: FT-1$\sim$FT-4 adopt full-parameter fine-tuning, while FT-5$\sim$FT-8 use parameter-efficient LoRA fine-tuning. Across setups ranging from purely synthetic data (7k) to a combination of synthetic and real data, and fully real data, both full-parameter and LoRA substantially improve ASR performance over zero-shot. Notably, LoRA achieves competitive results using far fewer trainable parameters. Furthermore, the amount of real data has a clear impact on recognition, but thanks to the strong representations of DMA-KWS, using only 100 real samples is sufficient to achieve performance comparable to 7k samples.

\begin{figure*}[!t]
\centerline{\includegraphics[width=\linewidth]{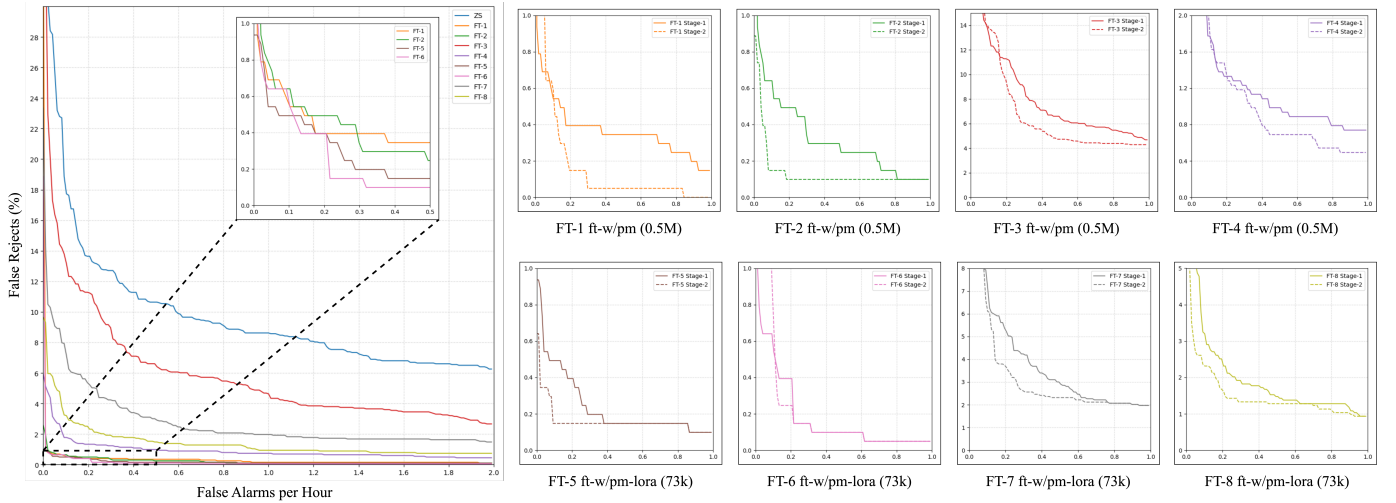}}
\caption{DET curve comparison of different Stage-1 fine-tuning strategies on the keyword ``OK Google''. Left: models using only Stage-1 fine-tuning (Table~\ref{tab:few-shot}); Right: phoneme-search models fine-tuned in Stage-2 with the same data and settings. Full-parameter fine-tuning involves 0.5M trainable parameters, while LoRA fine-tuning uses only 73k parameters.}
\label{fig.det}
\end{figure*}

\begin{figure}[!b]
\centerline{\includegraphics[width=0.85\linewidth]{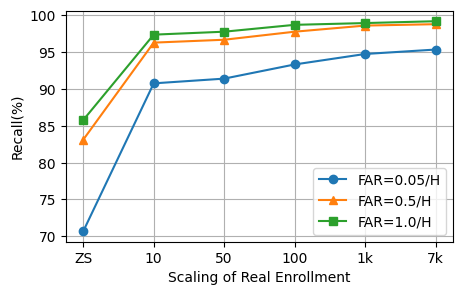}}
\caption{Recall performance on the keyword ``OK Google'' with Persian accent for Stage-2 only models (DMA-KWS(\Circled{2})) without CTC-Streaming, fine-tuned with varying amounts of real enrollment data.}
\label{fig.real}
\end{figure}

Next, Figure~\ref{fig.det} compares DET curves of different Stage-1 fine-tuning strategies on the keyword ``OK Google''. The left panel shows models using only Stage-1 fine-tuning (corresponding to Table~\ref{tab:few-shot}), while the right panel shows phoneme-search models fine-tuned in Stage-2 with the same data and settings. Full-parameter fine-tuning involves 0.5M trainable parameters, whereas LoRA fine-tuning uses only 73k parameters. The DET curves indicate that Stage-2 fine-tuning consistently improves detection performance across most strategies.

Finally, Figure~\ref{fig.real} shows the recall performance of Stage-2 only models (DMA-KWS(\Circled{2})) on the keyword ``OK Google'' with a Persian accent, fine-tuned with varying amounts of real enrollment data. A clear scaling effect is observed: the more real enrollment data used, the higher the recall. However, compared with the two-stage fine-tuning strategies shown in Figure~\ref{fig.det}, performance is slightly lower, further demonstrating the advantage of the two-stage approach.

\input{tabs/continual}

\reviseC{Beyond adaptation effectiveness, we further investigate whether specializing the model for a target keyword degrades its performance on generic tasks. To assess this robustness, we compare the frozen pre-trained model (Base) with the adapted versions (Full-tuning and LoRA) on generic datasets. As shown in Table~\ref{tab:continual_adap}, no significant performance degradation occurs across all generic tasks post-fine-tuning. Specifically, the P-WER and EER on non-target sets ($LS_{clean}$ and $LP_{H}$) remain stable, demonstrating that the model preserves its generalized discriminative power while adapting to new keywords.}

\subsection{\reviseB{Robustness and Error-type Analysis}}

\reviseB{
To further analyze the robustness of the proposed DMA-KWS, we conducted three controlled experiments targeting potential errors introduced by Stage 1, including posterior uncertainty, temporal localization shifts, and specific hard error cases. 
}

\input{tabs/phoneme-perturbation}

\reviseB{
We first simulate degradation in phoneme recognition by perturbing the CTC posterior probabilities. Specifically, we interpolate the posterior with a uniform distribution, i.e., $p=(1-\alpha)p+\alpha u$, where $\alpha$ controls the perturbation strength. As shown in Table~\ref{tab:phoneme_perturbation}, the recall of the CTC-streaming baseline degrades rapidly as $\alpha$ increases, while DMA-KWS remains significantly more stable. This demonstrates that the second-stage verification effectively compensates for the uncertainty in Stage 1 outputs.
}

\input{tabs/timestamp_perturbation}
\input{tabs/error_TYPE}

\reviseB{
Next, we evaluate robustness to temporal localization errors by perturbing the predicted timestamps from Stage 1. This simulates scenarios where the candidate segment is shifted and may contain additional preceding or trailing phonemes. As shown in Table~\ref{tab:timestamp_perturbation}, the performance remains largely stable under moderate perturbations (5\%-10\%), while more severe perturbations (20\%) lead to noticeable degradation. These results indicate that the second-stage matching is tolerant to moderate temporal misalignment and can handle additional phonetic interference.
}

\reviseB{
Finally, we analyzed three typical challenging cases: shared-prefix confusions, 1-phoneme substitution, and deletion errors. As shown in Table~\ref{tab:error_type_analysis}, the experimental data is derived from the filtered version of the LibriPhrase Hard set. The results indicate that the second stage consistently improves performance across all error types, with the most significant improvement observed in shared-prefix errors.
}

\begin{figure}[htb]
\centering
\includegraphics[width=0.9\linewidth]{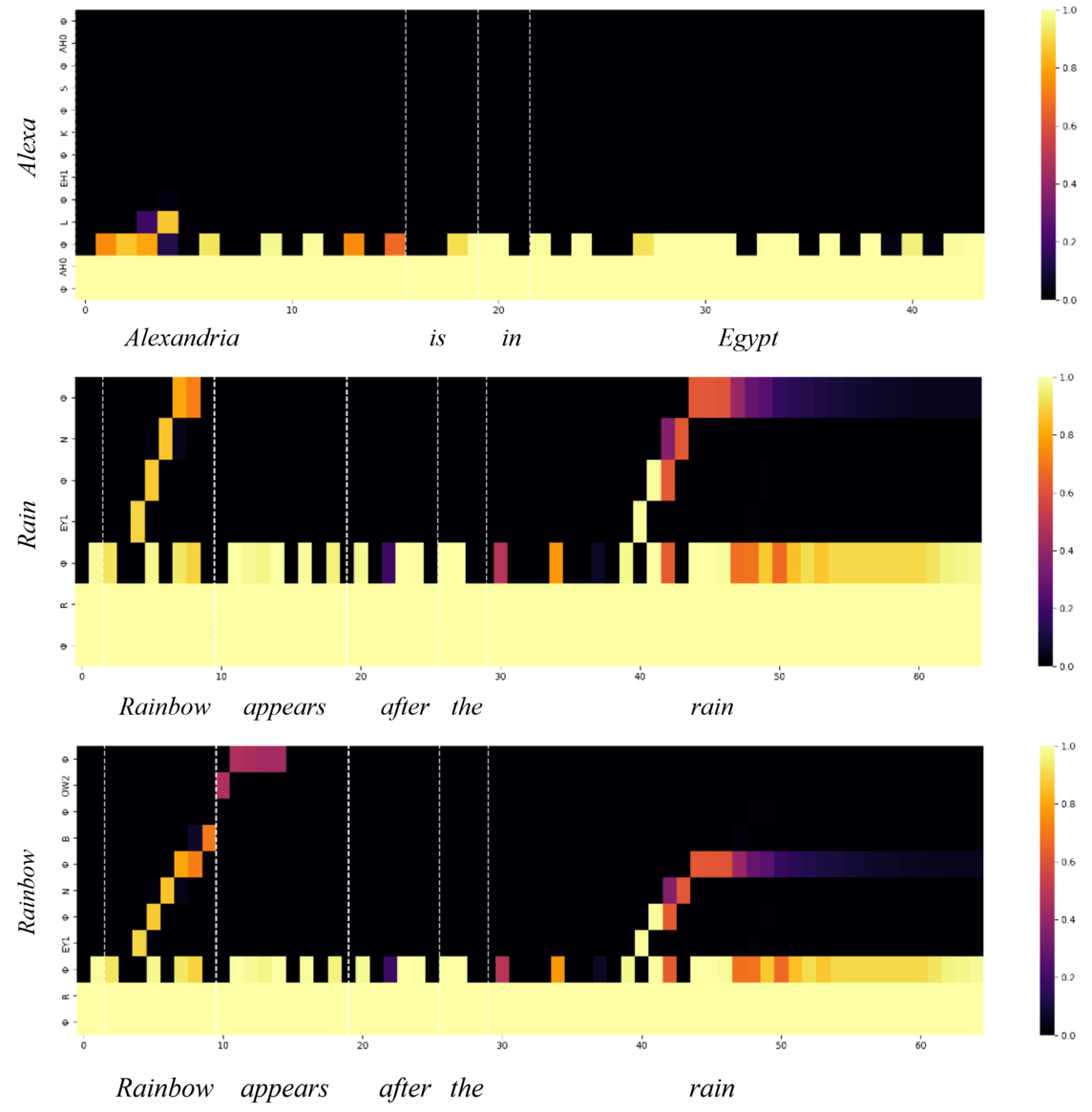}
\caption{\reviseC{Heatmaps of wake-up scores for prefix-sharing cases.}}
\label{fig:share-prefix}
\end{figure}

\reviseC{Additionally, we analyze the shared-prefix issue, which can be divided into two cases, as shown in Figure~\ref{fig:share-prefix}. The first case is phonetic discrepancy, where terms share the same prefix (e.g., "alexa" vs. "Alexander"), but their phoneme sequences differ\footnote{"Alexa" is /AH0 L EH1 K S AH0/, whereas "Alexander" is /AE2 L AH0 G Z AE1 N D R IY0 AH0/.}. CTC decoding helps prevent false triggers by recognizing these phonetic differences. The second case is phonetic overlap, where there is complete phonetic overlap between the words (e.g., "Rain" vs. "Rainbow"), causing the model to trigger the target word in the prefix region. To avoid relying on ASR or semantic recognition, we propose a negative-decoding strategy that suppresses the trigger of the shorter target word when the longer word has a higher confidence score.}

\subsection{Inference Efficiency}

This section evaluates the inference efficiency of the proposed DMA-KWS framework by measuring the total inference time on the LibriSpeech test set (10.75 h) under different hardware settings. As shown in Table \ref{tab:inference}, DMA-KWS introduces only a marginal overhead compared with the single-stage CTC-Streaming. The computational cost of the second-stage verification remains limited and does not hinder practical deployment on edge devices. In particular, DMA-KWS(\Circled{1}) achieves runtime nearly identical to CTC-Streaming, while DMA-KWS(\Circled{2}) trades a small increase in inference time for improved robustness. These results demonstrate that the proposed multi-stage framework preserves high inference efficiency while enhancing keyword verification capability.

\input{tabs/inference}

\begin{figure}[htp]
\centering
\includegraphics[width=\linewidth]{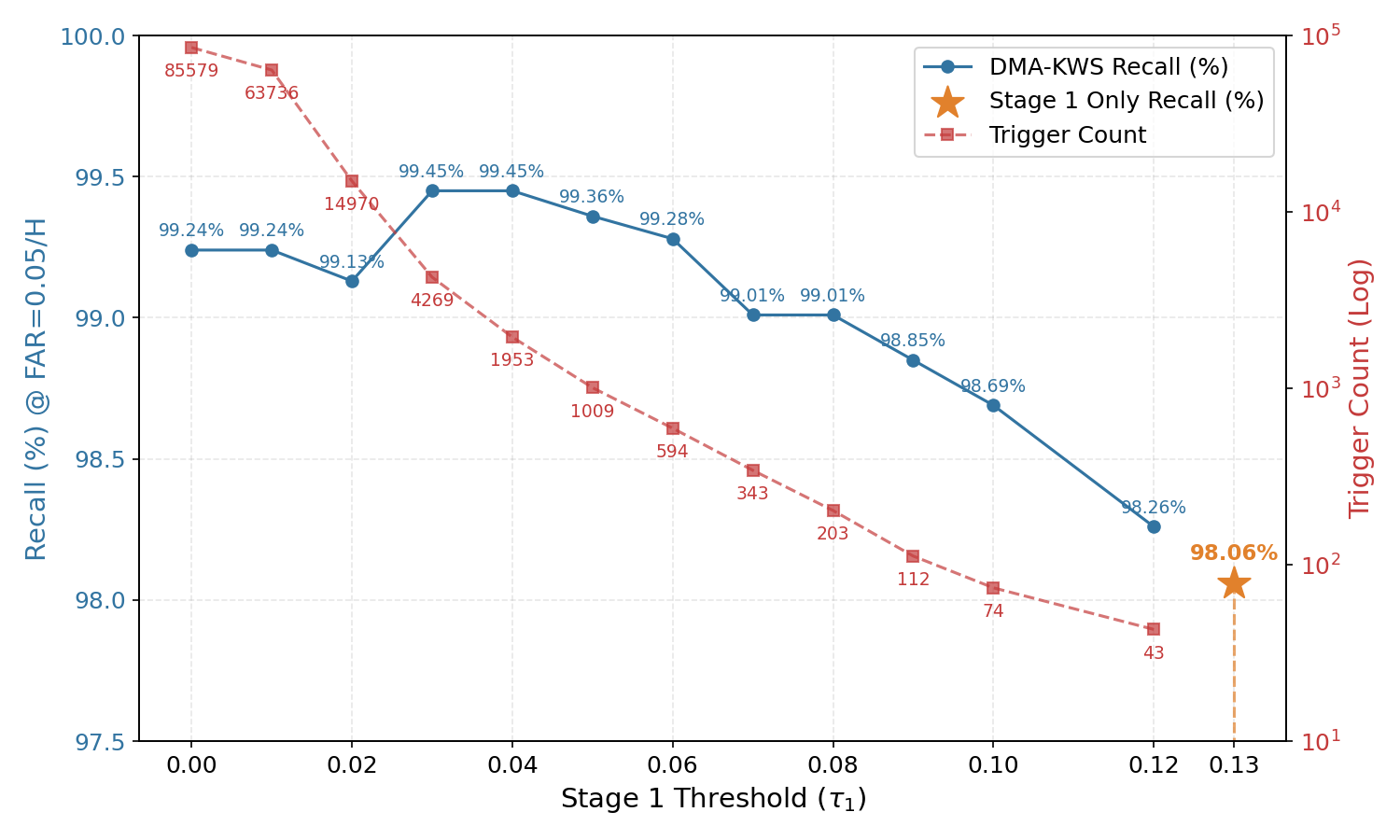}
\caption{\reviseC{Performance-Efficiency Trade-off of the cascaded DMA-KWS on HeySnips.}}
\label{fig:tradeoff}
\end{figure}

\reviseC{Figure~\ref{fig:tradeoff} shows the balance between the second-stage activation frequency and recall rate achieved by varying $\tau_1$ on the HeySnips dataset. Under the condition of a fixed false alarm rate (FAR = 0.05/H), the system achieves a peak recall of 99.45\% at $\tau_1 = 0.04$, triggering the second-stage verification only 1,953 times. Lowering $\tau_1$ below 0.04 leads to an exponential increase in redundant computations, with no significant gain in accuracy. Compared to the single-stage baseline (Recall 98.06\%), DMA-KWS achieves significant performance improvements with minimal additional computational overhead.}

%% file: tabs/si-libriphrase.tex
\begin{table*}[!t]
\caption{\reviseC{Experimental results of the proposed DMA-KWS model for speaker-independent keyword spotting, comparing DMA-KWS(\Circled{1}) (frozen audio encoder) and DMA-KWS(\Circled{2}) (trainable encoder). “–” indicates unavailable results. “*” denotes ASR-based pretraining, “$\dagger$” denotes phrase-based pretraining, and “$\ddagger$” denotes the use of synthetic data.}}
\centering
\resizebox{0.95\linewidth}{!}{
\begin{tabular}{@{}lcccccccccccccc@{}}
\toprule
\multirow{2}{*}{\textbf{Method}} & \multirow{2}{*}{\textbf{Venue}} & \multirow{2}{*}{\textbf{\# Params}} & \reviseC{\textbf{PT}} & \reviseC{\textbf{FT}} &  & \multicolumn{4}{c}{\textbf{AUROC (\%) ↑}} &  & \multicolumn{4}{c}{\textbf{EER (\%) ↓}} \\ \cmidrule(lr){7-10} \cmidrule(l){12-15} 
 &  &  & \reviseC{\textbf{(hrs)}} & \reviseC{\textbf{(hrs)}} &  & \textbf{LP}$_\textbf{H}$ & \textbf{LP}$_\textbf{E}$ & \textbf{GSC} & \textbf{Qcomm} &  & \textbf{LP}$_\textbf{H}$ & \textbf{LP}$_\textbf{E}$ & \textbf{GSC} & \textbf{Qcomm} \\ \midrule
Triplet \cite{baseline_triplet} & IS 19 & - & \reviseC{0} & \reviseC{460} &  & 54.88 & 63.53 & 71.48 & 66.44 &  & 44.36 & 32.75 & 35.60 & 38.72 \\
DONUT \cite{lugosch2018donut} & ICASSP 21 & - & \reviseC{0} & \reviseC{460} &  & 62.55 & 78.74 & 92.09 & 50.13 &  & 41.95 & 28.74 & 14.75 & 49.13 \\
CMCD \cite{baseline_cmcd} & IS 22 & 0.7M & \reviseC{0} & \reviseC{460} &  & 73.58 & 96.70 & 81.06 & 94.51 &  & 32.90 & 8.42 & 27.25 & 12.15 \\
EMKWS \cite{baseline_emkws} & IS 23 & 3.7M & \reviseC{460$\dagger$} & \reviseC{460} &  & 84.21 & 97.83 & - & - &  & 23.36 & 7.36 & - & - \\
PhonMatchNet \cite{baseline_phonmatchnet} & IS 23 & 0.7M & \reviseC{110,000*} & \reviseC{460} &  & 88.52 & 99.29 & 98.11 & 98.90 &  & 18.82 & 2.80 & 6.77 & 4.75 \\
CLAD \cite{baseline_clad} & ICASSP 24 & 3.6M & \reviseC{460*} & \reviseC{0} &  & 76.15 & 97.03 & - & - &  & 30.30 & 8.65 & - & - \\
CED \cite{baseline_ced} & ICASSP 24 & 3.6M & \reviseC{460$\dagger$} & \reviseC{460} &  & 92.70 & 99.84 & 93.94 & - &  & 14.40 & 1.70 & 13.45 & - \\
AdaKWS-Tiny \cite{baseline_adakws} & ICASSP 24 & 15M & \reviseC{\reviseC{1800$\dagger$}} & \reviseC{460} &  & 93.75 & 99.80 & - & - &  & 13.47 & 1.61 & - & - \\
AdaKWS-Small \cite{baseline_adakws} & ICASSP 24 & 109M & \reviseC{1800$\dagger$} &\reviseC{460}&  & 95.09 & 99.82 & - & - &  & 11.48 & 1.21 & - & - \\
CTCAT \cite{jin24d_interspeech} & IS 24 & 0.2M & \reviseC{0} &\reviseC{460}&  & 77.10 & 98.32 & - & - &  & 29.63 & 6.06 & - & - \\
MM-KWS$@$T \cite{ai24_interspeech} & IS 24 & 3.9M & \reviseC{0} & \reviseC{460+417$\ddagger$} &  & 95.36 & 99.94 & 98.69 & 99.71 &  & 10.41 & 0.82 & 5.52 & 2.31 \\
PLCL$@$T \cite{plcl} & ICASSP 25 & 40M &  \reviseC{680,000$\dagger$}  &\reviseC{460}&  & 95.56 & 99.95 & - & - &  & 9.96 & 1.21 & - & - \\
SLiCK \cite{SLiCK} & ICASSP 25 & 0.6M & \reviseC{460$\dagger$} &\reviseC{460}&  & 94.90 & 99.82 & - & - &  & 11.10 & 1.78 & - & - \\
ADML \cite{baseline_adml} & IS 25 & 1.8M & \reviseC{4600*} & \reviseC{0} &  & 88.71 & 99.86 & - & - &  & 20.09 & 1.33 & - & - \\
W-CTC \cite{kim25d_interspeech} & IS 25 & 3.6M & \reviseC{460$\dagger$} &\reviseC{460}&  & 95.93 & 99.95 & 98.89 & - &  & 10.21 & 0.91 & 4.64 & - \\ \midrule
MM-KWS$@$AT \cite{ai24_interspeech} & IS 24 & 3.9M & \reviseC{0} & \reviseC{460+417$\ddagger$} &  & 96.25 & 99.95 & 98.97 & 99.74 &  & 9.30 & 0.68 & 4.86 & 2.15 \\
PLCL$@$AT \cite{plcl} & ICASSP 25 & 40M &  \reviseC{680,000$\dagger$}  &\reviseC{460}&  & 96.59 & 99.97 & - & - &  & 8.47 & 0.57 & - & - \\ \midrule
\reviseC{DMA-KWS(\Circled{1})} &  & \reviseC{4.1M} & \reviseC{460$\dagger$} &\reviseC{460}&  & \reviseC{95.33} & \reviseC{99.96} & \reviseC{97.82} & \reviseC{99.78} &  & \reviseC{10.78} & \reviseC{0.72} & \reviseC{7.56} & \reviseC{2.17} \\
\reviseC{DMA-KWS(\Circled{2})} &  & \reviseC{4.1M} & \reviseC{460$\dagger$} &\reviseC{460}&  & \reviseC{97.03} & \reviseC{99.96} & \reviseC{98.30} & \reviseC{99.18} &  & \reviseC{7.97} & \reviseC{0.59} & \reviseC{5.95} & \reviseC{3.74} \\ \midrule
DMA-KWS(\Circled{1}) &  & 4.1M & \reviseC{1460$\dagger$} & \reviseC{1460} &  & 95.77 & 99.98 & 98.27 & 99.84 &  & 10.02 & 0.52 & 6.38 & 1.62 \\
DMA-KWS(\Circled{2}) &  & 4.1M & \reviseC{1460$\dagger$} & \reviseC{1460} &  & \textbf{97.85} & \textbf{99.98} & \textbf{99.21} & \textbf{99.90} &  & \textbf{6.13} & \textbf{0.45} & \textbf{3.93} & \textbf{1.52} \\ \bottomrule
\end{tabular}
}
\label{table:lp}
\end{table*}

%% file: tabs/data-scaling.tex
\begin{table}[!t]
\caption{Dual Data Scaling of the proposed DMA-KWS(\Circled{1} and \Circled{2}).}
\centering
\resizebox{\linewidth}{!}{
\begin{tabular}{@{}lcccccc@{}}
\toprule
\multirow{2}{*}{\textbf{Setting}} & \multicolumn{2}{c}{\textbf{P-WER (\%) ↓}}        & \multicolumn{2}{c}{\textbf{AUROC (\%) ↑}} & \multicolumn{2}{c}{\textbf{EER (\%) ↓}} \\ \cmidrule(l){2-7} 
                                  & \textbf{$\textbf{LS}_{\textbf{clean}}$} & \textbf{$\textbf{LS}_{\textbf{other}}$} & \textbf{$\textbf{LP}_{\textbf{H}}$}     & \textbf{$\textbf{LP}_{\textbf{E}}$}    & \textbf{$\textbf{LP}_{\textbf{H}}$}     & \textbf{$\textbf{LP}_{\textbf{E}}$}    \\ \midrule
\multicolumn{7}{l}{\textit{Stage1: LS-100; Stage 2: LP-100}}                 \\ \midrule
DMA-KWS(\Circled{1})     & 6.98 & 18.79 & 91.78 & 99.85 & 15.34 & 1.35 \\
DMA-KWS(\Circled{2})     & -    & -     & 93.10 & 99.88 & 13.71 & 1.14 \\ \midrule
\multicolumn{7}{l}{\textit{Stage1: LS-460; Stage 2: LP-460}}                 \\ \midrule
DMA-KWS(\Circled{1})     & \textbf{4.44} & \underline{13.39} & 95.33 & 99.96 & 10.78 & 0.72 \\
DMA-KWS(\Circled{2})     & -    & -     & \underline{97.03} & \underline{99.96} & \underline{7.97}  & \underline{0.59} \\ \midrule
\multicolumn{7}{l}{\textit{Stage1: LS-GS-1460; Stage 2: LP-GP-1460}}             \\ \midrule
DMA-KWS(\Circled{1}) & \underline{4.45} & \textbf{11.80} & 95.77 & 99.98 & 10.02 & 0.52 \\
DMA-KWS(\Circled{2})                     & -                   & -                   & \textbf{97.85}  & \textbf{99.98} & \textbf{6.13}   & \textbf{0.45}  \\ \bottomrule
\end{tabular}
}
\label{tab:dual_data_scaling}
\end{table}

%% file: tabs/anchor_scaling.tex
\begin{table}[!t]
\label{table:scaling2}
\caption{Effect of Anchor Scaling for DMA-KWS. The audio encoder, \textbf{pretrained on LS-460}, is frozen during training (DMA-KWS(\Circled{1})).}
\centering
\resizebox{0.9\linewidth}{!}{
\begin{tabular}{@{}lccccc@{}}
\toprule
\multirow{2}{*}{\textbf{Setting}} & \multirow{2}{*}{\textbf{\# Anchors}} & \multicolumn{2}{c}{\textbf{AUROC (\%) ↑}} & \multicolumn{2}{c}{\textbf{EER (\%) ↓}} \\ \cmidrule(l){3-6} 
             &      & \textbf{$\textbf{LP}_{\textbf{H}}$} & \textbf{$\textbf{LP}_{\textbf{E}}$} & \textbf{$\textbf{LP}_{\textbf{H}}$} & \textbf{$\textbf{LP}_{\textbf{E}}$} \\ \midrule
LP-100       & 12k  & 93.22       & 99.88       & 13.38       & 1.19        \\
LP-460-20k   & 20k  & 93.95       & 99.94       & 12.50       & 0.82        \\
LP-460-40k   & 40k  & 94.75       & 99.96       & 11.62       & 0.69        \\
LP-460       & 78k  & 95.33       & 99.96       & 10.78       & 0.72        \\
LP-GP-1460   & 155k & \textbf{95.45}       & \textbf{99.97}       & \textbf{10.65}       & \textbf{0.64}        \\ \bottomrule
\end{tabular}
}
\label{tab:stage2_ablation}
\end{table}

%% file: tabs/sd-libriphrase.tex
\begin{table*}[!t]
\caption{Experimental results of the proposed DMA-KWS for speaker-dependent keyword spotting, comparing DMA-KWS(\Circled{3}) (no extra parameters) and DMA-KWS(\Circled{4}) (no extra tokens).}
\centering
\resizebox{0.95\linewidth}{!}{
\begin{tabular}{@{}llllccccccccc@{}}
\toprule
\multirow{2}{*}{\textbf{Method}} & \multicolumn{2}{l}{\textbf{\# Params}} &  & \multicolumn{4}{c}{\textbf{AUROC (\%) ↑}} &  & \multicolumn{4}{c}{\textbf{EER (\%) ↓}} \\ \cmidrule(lr){2-3} \cmidrule(lr){5-8} \cmidrule(l){10-13} 
 & \textbf{Infer} & \textbf{Enroll} &  & $\textbf{LP}^\textbf{SD}_\textbf{H}$ & $\textbf{LP}^\textbf{SD}_\textbf{E}$ & \textbf{Qcomm}$^\textbf{SD}$ & \textbf{AudioMNIST} &  & $\textbf{LP}^\textbf{SD}_\textbf{H}$ & $\textbf{LP}^\textbf{SD}_\textbf{E}$ & \textbf{Qcomm}$^\textbf{SD}$ & \textbf{AudioMNIST} \\ \midrule
\multicolumn{13}{c}{\textit{\textbf{SI-Mode}}} \\ \midrule
MM-KWS$@$T \cite{ai24_interspeech} & 3.9M & 66.8M &  & 94.55 & 99.91 & 99.64 & 99.72 &  & 11.06 & 1.10 & 2.70 & 2.45 \\
DMA-KWS(\Circled{1})$@$T & 4.1M & \textbf{9K} &  & 95.37 & 99.97 & 99.84 & 99.65 &  & 10.73 & 0.52 & 1.62 & 2.97 \\
DMA-KWS(\Circled{2})$@$T & 4.1M & \textbf{9K} &  & 97.31 & 99.97 & 99.90 & 99.83 &  & 7.21 & 0.42 & 1.52 & 1.70 \\ \midrule
\multicolumn{13}{c}{\textit{\textbf{SD-Mode}}} \\\midrule
Whisper$@$A \cite{baseline_whisper} & 19.9M & 19.9M &  & 84.54 & 94.61 & 99.66 & 99.64 &  & 23.05 & 11.73 & 2.10 & 2.32 \\ 
WavLM$@$A \cite{chen2022wavlm} & 94.7M & 94.7M &  & 93.55 & 99.59 & \textbf{99.99} & \textbf{99.95} &  & 13.35 & 2.07 & \textbf{0.57} & \textbf{0.66} \\
MM-KWS$@$AT \cite{ai24_interspeech} & 3.9M & 398.6M &  & 96.68 & 99.96 & 99.91 & 99.28 &  & 8.27 & 0.60 & 1.36 & 3.82 \\ \midrule
DMA-KWS(\Circled{1})$@$A & 3.6M & 3.6M &  & 89.63 & 99.46 & 99.68 & 99.74 &  & 18.37 & 3.22 & 2.12 & 2.04 \\
DMA-KWS(\Circled{2})$@$A & 3.6M & 3.6M &  & 96.61 & 99.85 & 99.82 & 95.51 &  & 9.27 & 1.37 & 1.55 & \underline{0.90} \\
DMA-KWS(\Circled{3})$@$AT & 4.1M & 3.6M &  & 97.67 & 99.98 & 99.94 & \underline{99.93} &  & 6.70 & 0.38 & 1.04 & \underline{0.90} \\
DMA-KWS(\Circled{4})$@$AT & 4.1M & 3.6M &  & \textbf{97.70} & \textbf{99.98} & \underline{99.97} & 99.80 &  & \textbf{6.58} & \textbf{0.31} & \underline{0.88} & 1.67 \\ \bottomrule
\end{tabular}
}
\label{table:sd-lp}
\end{table*}

%% file: tabs/snips.tex
\begin{table}[htbp]
\caption{Zero-shot performance of DMA-KWS on Hey-Snips}
\centering
\resizebox{0.75\linewidth}{!}{
\begin{tabular}{@{}lcccc@{}}
\toprule
\multirow{2}{*}{\textbf{Method}} & \textbf{Training} & \multicolumn{3}{c}{\textbf{Recall (\%) @ FARs}} \\ \cmidrule(l){3-5} 
 & \textbf{Data} & \textbf{0.05} & \textbf{0.5} & \textbf{1.0} \\ \midrule
RIL-KWS \cite{zhang20u_interspeech} & \multirow{3}{*}{Official Snips} & - & 96.47 & 97.18 \\
WaveNet \cite{snips} &  & - & 99.88 & - \\
MDTC \cite{wekws} &  & - & 99.88 & 99.92 \\ \midrule
MDTC \cite{mfa-kws} & Pos. Snips & 89.52 & 98.85 & 99.29 \\
MFA-KWS \cite{mfa-kws} & + Equ. LS & 99.80 & 99.96 & 99.96 \\ \midrule
CTC-Streaming & \multirow{5}{*}{Zero-shot} & 98.06 & 98.89 & 98.97 \\
DMA-KWS(\Circled{1}) &  & 98.66 & 99.28 & 99.45 \\
DMA-KWS(\Circled{2}) &  & 99.45 & 99.76 & 99.80 \\
DMA-KWS(\Circled{3}) &  & 99.64 & 99.84 & 99.84 \\
DMA-KWS(\Circled{4}) &  & 99.72 & 99.84 & 99.84 \\ \bottomrule
\end{tabular}
}
\label{tab:snips_results}
\end{table}

\begin{table}[htbp]
\caption{Recall comparison at FAR = 0.05/h for zero-shot performance on additional custom wake words}
\centering
\resizebox{0.95\linewidth}{!}{
\begin{tabular}{@{}lccccl@{}}
\toprule
\textbf{Method} & \textbf{Lumina} & \textbf{Galaxy} & \textbf{Snapdragon} & \textbf{Android} & \textbf{Avg.} \\ \midrule
CTC-Streaming & 82.88 & 89.72 & 96.15 & 86.72 & 88.87 \\
DMA-KWS(\Circled{1}) & 90.66 & 94.20 & 98.81 & 94.32 & 94.50 \\
DMA-KWS(\Circled{2}) & 93.59 & 98.80 & 98.81 & 96.25 & 96.86 \\
DMA-KWS(\Circled{3}) & 94.23 & 99.56 & 99.72 & 98.53 & 98.01 \\
DMA-KWS(\Circled{4}) & 95.97 & 99.67 & 99.72 & 99.45 & 98.70 \\ \bottomrule
\end{tabular}
}
\label{tab:customs_results}

\end{table}

%% file: tabs/few-shot-kws.tex
\begin{table}[!b]
\caption{Stage-1 ASR fine-tuning on DeepMine. ZS denotes zero-shot; FT-1$\sim$4 are full-parameter fine-tuning, and FT-5$\sim$8 use LoRA. \#Real / \#Syn indicate the number of real / synthetic samples per target word. T1 corresponds to ``OK Google''.}
\centering
\resizebox{\linewidth}{!}{
\begin{tabular}{@{}ccccccccc@{}}
\toprule
\multirow{2}{*}{\textbf{Method}} & \multirow{2}{*}{\textbf{\# \textbf{Real}}} & \multirow{2}{*}{\textbf{\# \textbf{Syn}}} & \multirow{2}{*}{\textbf{LORA}} & \multicolumn{5}{c}{\textbf{P-WER (\%) ↓}} \\ \cmidrule(l){5-9} 
 &  &  &  & \textbf{T1} & \textbf{T2} & \textbf{T3} & \textbf{T4} & \textbf{T5} \\ \midrule
ZS & - & - & - & 49.56 & 36.67 & 30.82 & 27.06 & 26.77 \\ \midrule
FT-1 & 100 & - & \multirow{2}{*}{$\times$} & 0.45 & 0.54 & 0.99 & 0.42 & 0.77 \\
FT-2 & 7k & - &  & 0.39 & 0.26 & 0.59 & 0.32 & 0.40 \\
FT-3 & - & 7k & \multirow{2}{*}{(3.6M)} & 14.59 & 18.64 & 15.17 & 7.77 & 10.99 \\
FT-4 & 10 & 7k &  & 2.79 & 5.56 & 5.17 & 2.00 & 4.67 \\ \midrule
FT-5 & 100 & - & \multirow{2}{*}{$\checkmark$} & 0.70 & 1.09 & 1.68 & 0.85 & 1.76 \\
FT-6 & 7k & - &  & 0.56 & 0.59 & 1.24 & 0.72 & 1.10 \\
FT-7 & - & 7k & \multirow{2}{*}{(114k)} & 16.08 & 20.99 & 18.87 & 12.30 & 13.62 \\
FT-8 & 10 & 7k &  & 10.94 & 9.74 & 11.53 & 5.11 & 7.52 \\ \bottomrule
\end{tabular}
}
\label{tab:few-shot}
\end{table}

%% file: tabs/continual.tex
\begin{table}[htp]
\caption{\reviseC{Comparison of system performance on generic tasks before and after targeted adaptation.}}
\centering
\reviseC{
\begin{tabular}{@{}lcccc@{}}
\toprule
\multirow{2}{*}{\textbf{Method}} & \multicolumn{2}{c}{\textbf{P-WER (\%) ↓}} & \multicolumn{2}{c}{\textbf{AUROC (\%) ↑ / EER (\%) ↓}} \\ \cmidrule(l){2-5} 
                                 & $\textbf{LS}_\textbf{clean}$   & $\textbf{LS}_\textbf{other}$   & $\textbf{LP}_\textbf{H}$             & $\textbf{LP}_\textbf{E}$            \\ \midrule
Base                               & 4.45               & 11.80               & 95.77 / 10.02            & 99.98 / 0.52            \\ \midrule
Full-tuning (4.1M)                            & 4.59               & 12.65              & 95.22 / 10.76            & 99.98 / 0.52            \\ \midrule
LoRA (187k)                            & 4.68               & 12.71              & 95.50 / 10.27            & 99.97 / 0.59            \\ \bottomrule
\end{tabular}
}
\label{tab:continual_adap}
\end{table}

%% file: tabs/phoneme-perturbation.tex
\begin{table}[htp]
\caption{\reviseB{Robustness under CTC posterior perturbation on Hey-Snips.}}
\centering
\reviseB{
\begin{tabular}{@{}lcccc@{}}
\toprule
\multirow{2}{*}{\textbf{Method}} & \textbf{Phoneme}      & \multicolumn{3}{c}{\textbf{Recall(\%) @ FARs}} \\ \cmidrule(l){3-5} 
                                 & \textbf{Perturbation} & \textbf{0.05}   & \textbf{0.5}   & \textbf{1}  \\ \midrule
\multirow{3}{*}{CTC-Streaming}   & 0\%                   & 98.06           & 98.89          & 98.97       \\
                                 & 10\%                  & 97.55           & 98.30          & 98.58       \\
                                 & 20\%                  & 96.48           & 97.82          & 98.14       \\ \midrule
\multirow{3}{*}{DMA-KWS(\Circled{1})}      & 0\%                   & 98.66           & 99.28          & 99.45       \\
                                 & 10\%                  & 97.55           & 98.93          & 99.24       \\
                                 & 20\%                  & 97.31           & 98.45          & 98.77       \\ \midrule
\multirow{3}{*}{DMA-KWS(\Circled{2})}      & 0\%                   & 99.45           & 99.76          & 99.80       \\
                                 & 10\%                  & 98.77           & 99.48          & 99.52       \\
                                 & 20\%                  & 98.26           & 99.25          & 99.28       \\ \bottomrule
\end{tabular}
}
\label{tab:phoneme_perturbation}
\end{table}

%% file: tabs/timestamp_perturbation.tex
\begin{table}[htp]
\caption{\reviseB{Robustness under timestamp perturbation on LibriPhrase.}}
\centering
\reviseB{
\begin{tabular}{@{}lccccc@{}}
\toprule
\multirow{2}{*}{\textbf{Method}} & \textbf{Timestamp}    & \multicolumn{2}{c}{\textbf{AUROC (\%) ↑}} & \multicolumn{2}{c}{\textbf{EER (\%) ↓}} \\ \cmidrule(l){3-6} 
                                 & \textbf{Perturbation} & \textbf{\textbf{LP}$_\textbf{H}$ }    & \textbf{\textbf{LP}$_\textbf{E}$ }   & \textbf{\textbf{LP}$_\textbf{H}$ }    & \textbf{\textbf{LP}$_\textbf{E}$ }   \\ \midrule
\multirow{4}{*}{DMA-KWS(\Circled{1})}      & 0\%                   & 95.77           & 99.98          & 10.02           & 0.52           \\
                                 & 5\%                   & 96.02           & 99.97          & 9.63            & 0.50           \\
                                 & 10\%                  & 95.84           & 99.97          & 9.85            & 0.63           \\
                                 & 20\%                  & 93.83           & 99.90          & 13.09           & 1.31           \\ \midrule
\multirow{4}{*}{DMA-KWS(\Circled{2})}      & 0\%                   & 97.85           & 99.98          & 6.13            & 0.45           \\
                                 & 5\%                   & 97.63           & 99.97          & 6.59            & 0.56           \\
                                 & 10\%                  & 96.73           & 99.93          & 8.47            & 0.86           \\
                                 & 20\%                  & 94.20           & 99.63          & 11.95           & 2.64           \\ \bottomrule
\end{tabular}
}
\label{tab:timestamp_perturbation}
\end{table}

%% file: tabs/error_TYPE.tex
\begin{table}[hbp]
\caption{\reviseB{Robustness analysis across different error types on the LibriPhrase hard subset (AUROC / EER, \%).}}
\centering
\resizebox{\linewidth}{!}{
\reviseB{
\begin{tabular}{@{}lccc@{}} 
\toprule
\multirow{2}{*}{\textbf{Method}} & \multicolumn{3}{c}{\textbf{Error Type}} \\ \cmidrule(l){2-4} 
                                 & \textbf{Shared-prefix} & \textbf{1-phoneme sub.} & \textbf{Deletion errors} \\ \midrule
Stage 1                    & 54.87 / 47.26          & 79.23 / 26.90           & 97.74 / 5.57             \\ \midrule
Stage 2                & 96.63 / 9.19           & 96.90 / 7.85            & 98.95 / 4.26             \\ \bottomrule
\end{tabular}
}
}
\label{tab:error_type_analysis}
\end{table}

%% file: tabs/inference.tex
\begin{table}[htp]
\caption{Total inference time of the proposed DMA-KWS on the LibriSpeech test set (\textbf{10.75 h}) under different hardware settings, evaluated with a single process.}
\centering
\resizebox{\linewidth}{!}{
\begin{tabular}{@{}llccc@{}}
\toprule
\multicolumn{2}{l}{\multirow{2}{*}{\textbf{Method}}} & RTX 4090 D & Intel i9-14900k & Macmini M2 \\
\multicolumn{2}{l}{} & GPU & CPU & CPU \\ \midrule
\multirow{2}{*}{\textbf{Stage1}} & CTC & 54s & 185s & 263s \\
 & Decoding & $\times$ & 409s & 496s \\ \midrule
\multirow{2}{*}{\textbf{Stage2}} & \Circled{1} (Enc-Clips) & 3s & 10s & 10s \\
 & \Circled{2} (Wav-Clips) & 18s & 70s & 88s \\ \midrule
\multirow{3}{*}{\textbf{Total}} & CTC-Streaming & 463s & 594s & 759s \\
 & DMA-KWS(\Circled{1}) & 466s & 604s & 769s \\
 & DMA-KWS(\Circled{2}) & 481s & 664s & 847s \\ \bottomrule
\end{tabular}
}
\label{tab:inference}
\end{table}

%% file: sections/6_conclusion.tex
\section{conclusion}
We propose DMA-KWS, an efficient and robust framework for user-defined keyword spotting. It integrates a coarse-to-fine dual-stage matching pipeline, combining CTC-based streaming phoneme search for candidate localization with QbyT-based phoneme matching for fine-grained verification. Multi-modal enrollment fuses user-specific speech with text embeddings to improve recognition for registered users, while a parameter-efficient continual adaptation mechanism enables rapid fine-tuning with minimal synthetic and real data. Extensive experiments show that DMA-KWS achieves state-of-the-art performance on multiple datasets, provides strong zero-shot capability, effectively distinguishes confusable keywords, and supports fast adaptation for newly enrolled keywords with very few updated parameters, making it well-suited for on-device deployment.